\documentclass{nature}

\usepackage{xspace}
\usepackage{color}
\usepackage{graphics}
\usepackage{epsfig}
\usepackage{upgreek}
\usepackage{amssymb}
\usepackage{amsmath}
\usepackage{siunitx}
\usepackage{booktabs}
\usepackage{mathrsfs}
\usepackage{url}
\usepackage{caption}

\usepackage{lineno}

\usepackage{xcolor}
\definecolor{orange}{HTML}{FFA500}
\definecolor{cyan}{HTML}{48D1CC}
\definecolor{purple}{HTML}{8D32C1}
\definecolor{green}{HTML}{008000}

\usepackage{caption}\usepackage[margin=1.3 in]{geometry}

\def\lapp{\ifmmode\stackrel{<}{_{\sim}}\else$\stackrel{<}{_{\sim}}$\fi}
\def\gapp{\ifmmode\stackrel{>}{_{\sim}}\else$\stackrel{>}{_{\sim}}$\fi}

\newcommand{\mr}{\mathrm}

\newcommand{\PC}{\textcolor{black}}
\newcommand{\vk}{\textcolor{black}}

\newcommand{\AJ}{\textcolor{black}}

\newcommand{\bca}{\textcolor{black}}

\newcommand{\nbursts}{38}

\title{Periodic activity from a fast radio burst source}

\author{
The CHIME/FRB Collaboration,
Amiri, M.$^{1}$, \allowbreak
Andersen, B.~C.$^{2,3}$, \allowbreak
Bandura, K.~M.$^{4,5}$, \allowbreak
Bhardwaj, M.$^{2,3}$, \allowbreak
Boyle, P.~J.$^{2,3}$, \allowbreak
Brar, C.$^{2,3}$, \allowbreak
Chawla, P.$^{2,3}$, \allowbreak
Chen, T.$^{6}$, \allowbreak
Cliche, J.~F.$^{2,3}$, \allowbreak
Cubranic, D.$^{1}$, \allowbreak
Deng, M.$^{1}$, \allowbreak
Denman, N.~T.$^{7}$, \allowbreak
Dobbs, M.$^{2,3}$, \allowbreak
Dong, F.~Q.$^{1}$, \allowbreak
Fandino, M.$^{1}$, \allowbreak
Fonseca, E.$^{2,3}$, \allowbreak
Gaensler, B.~M.$^{8,9}$, \allowbreak
Giri, U.$^{10,11}$, \allowbreak
Good, D.~C.$^{1}$, \allowbreak
Halpern, M.$^{1}$, \allowbreak
Hessels, J.~W.~T.$^{12,13}$, \allowbreak
Hill, A.~S.$^{14,15}$, \allowbreak
H\"ofer, C.$^{1}$, \allowbreak
Josephy, A.$^{2,3}$, \allowbreak
Kania, J.~W.$^{16}$, \allowbreak
Karuppusamy, R.$^{17}$, \allowbreak
Kaspi, V.~M.$^{2,3}$, \allowbreak
Keimpema, A.$^{18}$, \allowbreak
Kirsten, F.$^{19}$, \allowbreak
Landecker, T.~L.$^{15}$, \allowbreak
Lang, D.~A.$^{10,11}$, \allowbreak
Leung, C.$^{6,20}$, \allowbreak
Li, D.~Z.$^{21,22,17,8,*}$, \allowbreak
Lin, H.-H.$^{21,17}$, \allowbreak
Marcote, B.$^{18}$, \allowbreak
Masui, K.~W.$^{6,20}$, \allowbreak
Mckinven, R.$^{8,9}$, \allowbreak
Mena-Parra, J.$^{6}$, \allowbreak
Merryfield, M.$^{2,3}$, \allowbreak
Michilli, D.$^{2,3}$, \allowbreak
Milutinovic, N.$^{1,15}$, \allowbreak
Mirhosseini, A.$^{1}$, \allowbreak
Naidu, A.$^{2,3}$, \allowbreak
Newburgh, L.~B.$^{23}$, \allowbreak
Ng, C.$^{8}$, \allowbreak
Nimmo, K.$^{12,13}$, \allowbreak
Paragi, Z.$^{18}$, \allowbreak
Patel, C.$^{8,2,3}$, \allowbreak
Pen, U.-L.$^{21,8,24,10,17}$, \allowbreak
Pinsonneault-Marotte, T.$^{1}$, \allowbreak
Pleunis, Z.$^{2,3}$, \allowbreak
Rafiei-Ravandi, M.$^{10}$, \allowbreak
Rahman, M.$^{8}$, \allowbreak
Ransom, S.~M.$^{25}$, \allowbreak
Renard, A.$^{8}$, \allowbreak
Sanghavi, P.$^{4,5}$, \allowbreak
Scholz, P.$^{8,15}$, \allowbreak
Shaw, J.~R.$^{1}$, \allowbreak
Shin, K.$^{6,20}$, \allowbreak
Siegel, S.~R.$^{2,3}$, \allowbreak
Singh, S.$^{2,3}$, \allowbreak
Smegal, R.~J.$^{1}$, \allowbreak
Smith, K.~M.$^{10}$, \allowbreak
Stairs, I.~H.$^{1}$, \allowbreak
Tendulkar, S.~P.$^{2,3}$, \allowbreak
Tretyakov, I.$^{8,22}$, \allowbreak
Vanderlinde, K.$^{8,9}$, \allowbreak
Wang, H.$^{6,20}$, \allowbreak
Wang, X.$^{26}$, \allowbreak
Wulf, D.$^{2,3}$, \allowbreak
Yadav, P.$^{1}$, \allowbreak
Zwaniga, A.~V.$^{2,3}$ \allowbreak
}
\newcommand{\affils}{
\begin{affiliations}
\item{Department of Physics and Astronomy, University of British Columbia, 6224 Agricultural Road, Vancouver, BC V6T 1Z1, Canada}
\item{Department of Physics, McGill University, 3600 rue University, Montr\'eal, QC H3A 2T8, Canada}
\item{McGill Space Institute, McGill University, 3550 rue University, Montr\'eal, QC H3A 2A7, Canada}
\item{CSEE, West Virginia University, Morgantown, WV 26505, USA}
\item{Center for Gravitational Waves and Cosmology, West Virginia University, Morgantown, WV 26505, USA}
\item{MIT Kavli Institute for Astrophysics and Space Research, Massachusetts Institute of Technology, 77 Massachusetts Ave, Cambridge, MA 02139, USA}
\item{Central Development Laboratory, National Radio Astronomy Observatory, 1180 Boxwood Estate Road, Charlottesville VA 22903, USA}
\item{Dunlap Institute for Astronomy \& Astrophysics, University of Toronto, 50 St.~George Street, Toronto, ON M5S 3H4, Canada}
\item{David A.~Dunlap Department of Astronomy \& Astrophysics, University of Toronto, 50 St.~George Street, Toronto, ON M5S 3H4, Canada}
\item{Perimeter Institute for Theoretical Physics, 31 Caroline Street N, Waterloo ON N2L 2Y5, Canada}
\item{Department of Physics and Astronomy, University of Waterloo, Waterloo, ON N2L 3G1, Canada}
\item{ASTRON, Netherlands Institute for Radio Astronomy, Oude Hoogeveensedijk 4, 7991 PD Dwingeloo, The Netherlands}
\item{Anton Pannekoek Institute for Astronomy, University of Amsterdam, Science Park 904, 1098 XH Amsterdam, The Netherlands}
\item{Department of Computer Science, Math, Physics, and Statistics, University of British Columbia, 3187 University Way, Kelowna, BC V1V 1V7, Canada}
\item{Dominion Radio Astrophysical Observatory, Herzberg Research Centre for Astronomy and Astrophysics, National Research Council Canada, PO Box 248, Penticton, BC V2A 6J9, Canada}
\item{Department of Physics and Astronomy, West Viriginia University, Morgantown, WV 26505, USA}
\item{Max Planck Institute for Radio Astronomy, Auf dem Huegel 69, Bonn, 53121, Germany}
\item{Joint Institute for VLBI ERIC (JIVE), Oude Hoogeveensedijk 4, 7991 PD Dwingeloo, The Netherlands}
\item{Department of Space, Earth and Environment, Chalmers University of Technology, Onsala Space Observatory, 439 92, Onsala, Sweden}
\item{Department of Physics, Massachusetts Institute of Technology, 77 Massachusetts Ave, Cambridge, MA 02139, USA}
\item{Canadian Institute for Theoretical Astrophysics, 60 St.~George Street, Toronto, ON M5S 3H8, Canada}
\item{Department of Physics, University of Toronto, 60 St.~George Street, Toronto, ON M5S 1A7, Canada}
\item{Department of Physics, Yale University, New Haven, CT 06520, USA}
\item{Canadian Institute for Advanced Research, CIFAR Program in Gravitation and Cosmology, Toronto, ON M5G 1Z8, Canada}
\item{National Radio Astronomy Observatory, 520 Edgemont Road, Charlottesville, VA 22903, USA}
\item{School of Physics and Astronomy, Sun Yat-sen University, 2 Daxue Road, Zhuhai, China}
\end{affiliations}
}

\begin{document}

\maketitle

\blfootnote{
* Corresponding Author\\
%$\ddagger$ Members of the CHIME/FRB collaboration 
\affils
}
%\clearpage

\begin{abstract}
Fast radio bursts (FRBs) are bright, millisecond-duration radio transients originating from extragalactic
distances\cite{lbm+07}. Their origin is unknown. Some FRB sources emit repeat bursts, ruling out cataclysmic origins for those events\cite{ssh+16a,abb+19b,abb+19c}. Despite searches for periodicity in repeat burst arrival times on time scales from milliseconds to many days\cite{ssh+16a, ssh+16b,zgf+18,gms+19}, these bursts have hitherto been observed to appear sporadically, and though clustered\cite{18Oppermann}, without a regular pattern. Here we report the detection of a $16.35\pm0.15$ day periodicity (or possibly a higher-frequency alias of that periodicity)
from a repeating FRB 180916.J0158+65 detected by the Canadian Hydrogen Intensity Mapping Experiment Fast Radio Burst Project (CHIME/FRB) \cite{abb+18,abb+19c}. In \nbursts bursts recorded from September 16th, 2018 through February 4th, 2020, we find that all bursts arrive in a 5-day phase window, 
and 50\% of the bursts arrive in a 0.6-day phase window.
Our results suggest a mechanism for periodic modulation either of the burst emission itself,
or through external amplification or absorption, and disfavour 
models invoking purely sporadic processes.
\end{abstract}

Last year the CHIME/FRB collaboration reported the discovery of eight new repeating FRB sources\cite{abb+19c}, including FRB 180916.J0158+65, which was recently localized to a star-forming
region in a nearby massive spiral galaxy
at redshift 0.0337$\pm$0.0002\cite{mnh+20}. 
From September 2018 to February 2020, CHIME/FRB has detected a total of \nbursts bursts from FRB 180916.J0158+65. It is the most active source in the published CHIME/FRB sample.
The barycentric arrival times for the \nbursts bursts (including those have been published before) from FRB 180916.J0158+65, corrected for delays from pulse dispersion, are listed in Extended Data Table~\ref{tab:bursts}. 

 To search for periodicity, the burst arrival times (spanning a 500-day time range) were folded with different periods from 1.5 to 100 days (see Methods), with a Pearson's $\chi^2$ test applied to
 resulting profiles created with different numbers of phase bins\cite{Leahy1983}. 
 A reduced $\chi^2 \gg 1$ with respect to a uniform distribution indicates a periodicity unlikely to arise by chance. Furthermore, to account for the possible non-Poissonian statistics of the bursts\cite{18Oppermann}, we have applied the search with different weighting schemes that consider clustered bursts of different time range to be correlated events (see Methods).
 
\begin{figure}[htbp]
\begin{center}
\includegraphics[width=0.5\textwidth]{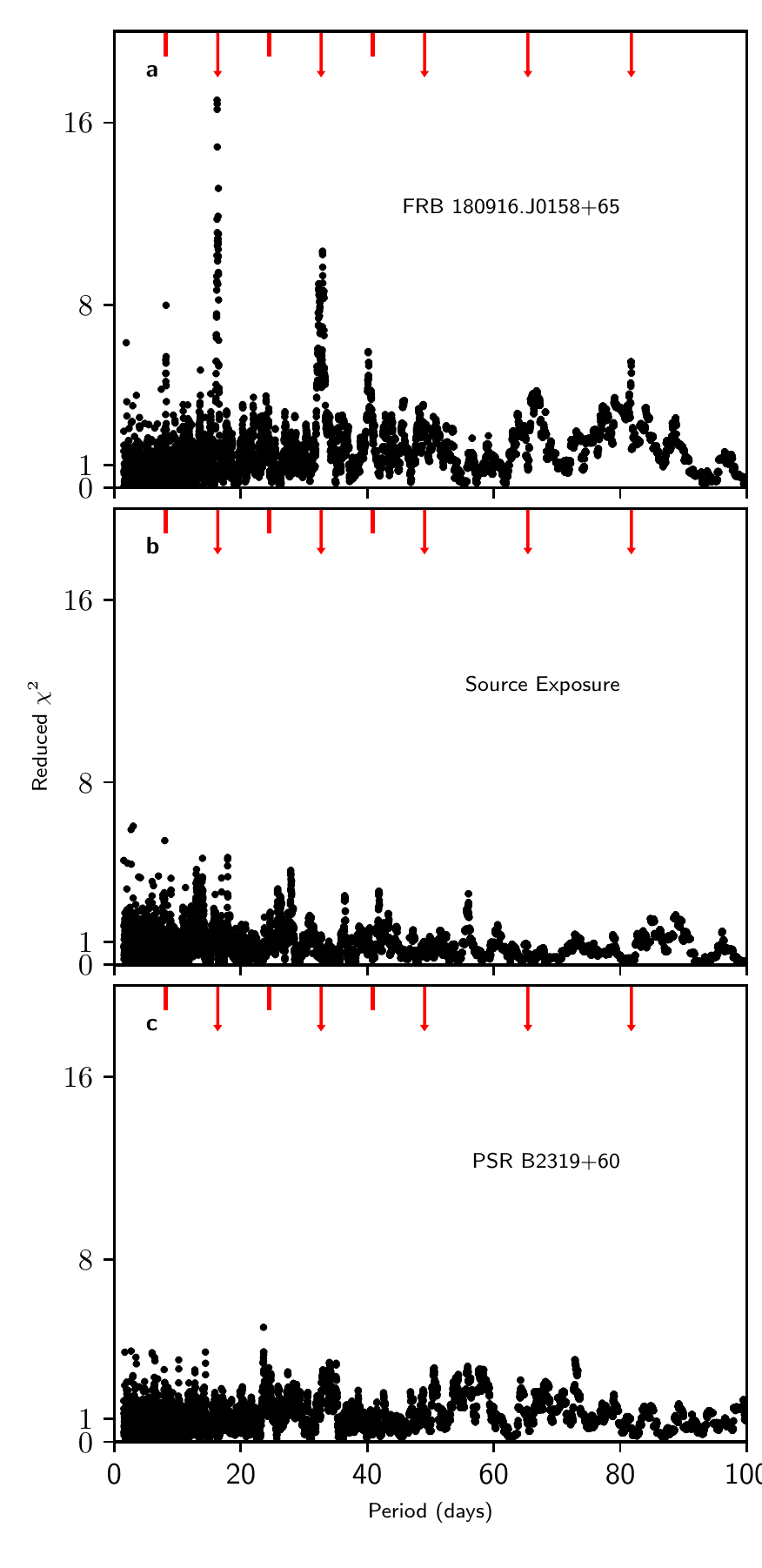}
\end{center}
%\end{figure}

%\clearpage
%\addtocounter{figure}{-1}
%\begin{figure} [t!]
\begin{center}
\caption{
\textbf{Periodograms of FRB 180916.J0158+65 and control samples.}
\textbf{a}: the reduced $\chi^2$ with respect to a uniform distribution of burst arrival times for different folding periods for FRB 180916.J0158+65 detected by CHIME/FRB. Only bursts separated by a sidereal day are considered independent in this approach. Details of the calculation of $\chi^2$ and other approaches are presented in Methods. 
\textbf{b}: the periodogram of mock burst arrival times randomly sampled according to the daily exposure to FRB 180916.J0158+65 within the FWHM of the telescope’s synthesized beams at 600 MHz.
\textbf{c}: the periodogram of randomly selected pulses of Galactic radio pulsar B2319+60 detected by the same instrument and software. 
The arrows indicate the first 5 
subharmonics of the 16.35-day periodicity, while the vertical lines mark the 
harmonics of 1/2 period. 
}
\label{fig:periodogram}
\end{center}
\end{figure}

Searches with different weightings return periodograms of similar shape and have the same primary peak with significance varying between $4.5-10\sigma$. 
As an example, the reduced $\chi^2$ versus period with 5 phase bins, using a weighting that counts only
active days instead of individual events, is shown in  Figure~\ref{fig:periodogram}a. 
A distinct peak is detected at $16.35\pm0.15$ days, with a probability of chance occurrence $\sim10^{-7}$ (equivalent to 5$\sigma$),   accounting for the number of independent periods searched. The other $\chi^2$ peaks correspond to harmonics and subharmonics of the period, with the one at 32.7 days being the next most prominent.
As a check, the same procedure was run on two types of control samples: (1) a mock data set, consisting of burst arrival times randomly sampled according to the daily exposure of FRB 180916.J0158+65 with the full width at half maximum (FWHM) of the telescope’s synthesized beams at 600 MHz (Figure~\ref{fig:periodogram}b) and (2) randomly selected pulses from Galactic radio pulsars of similar declination detected by CHIME/FRB (Figure~\ref{fig:periodogram}c) which have the same limited daily exposure and long-term sensitivity changes as the data for FRB 180916.J0158+65. The 16.35-day periodicity is absent in the control samples, and with $10^6$ sets of control samples of each type, no other periods 
have reached the level of significance of the 16.35-day period for FRB 180916.J0158+65 by chance coincidence. 

\begin{figure}[htbp]
  \includegraphics[width=1.1\linewidth,scale=1.2]{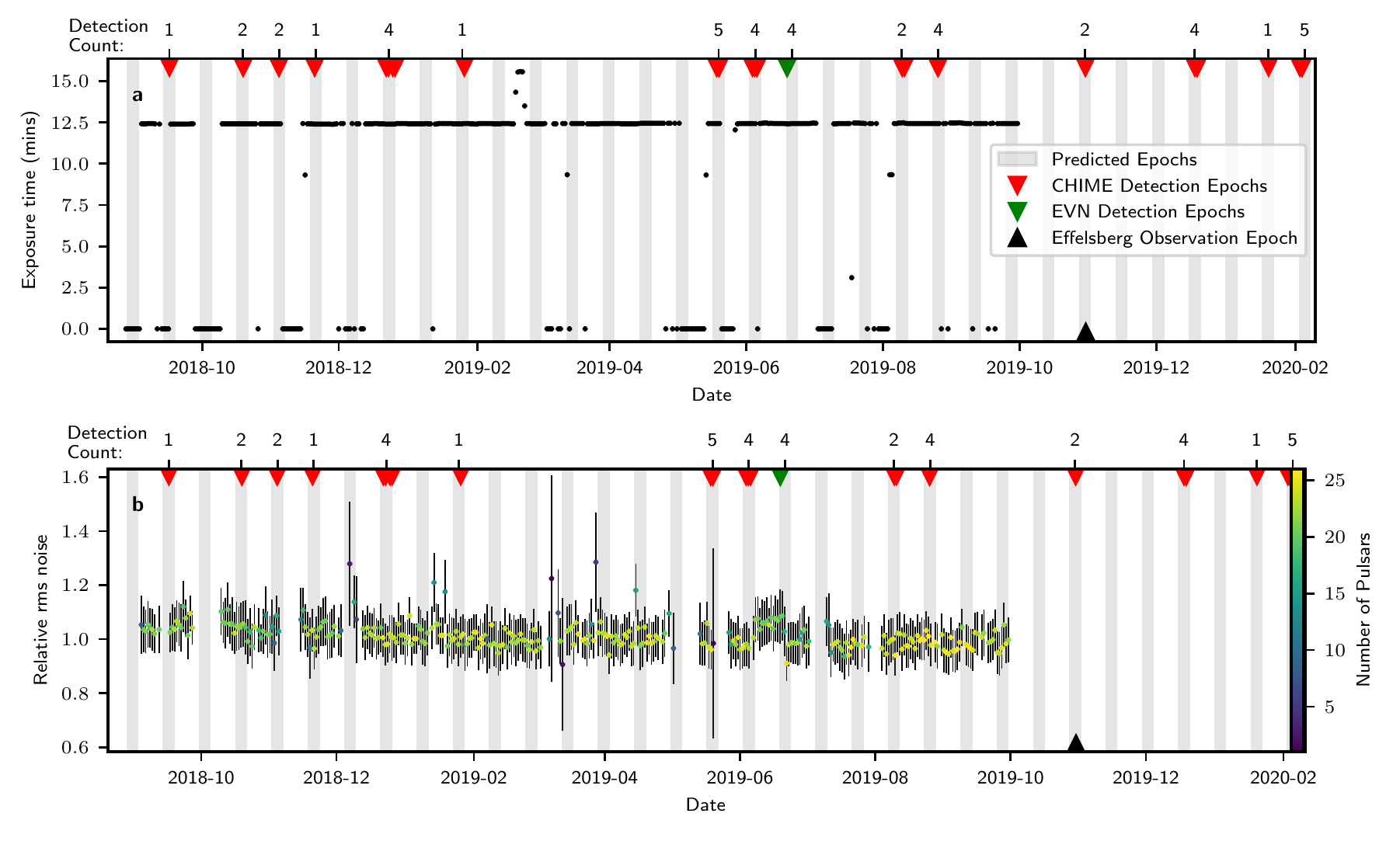}
%\end{figure}

%\clearpage
%\addtocounter{figure}{-1}
%\begin{figure}[t!]
  	  \caption{\textbf{Timeline of CHIME/FRB's daily exposure to FRB 180916.J0158+65.}
	  \textbf{a}: The exposure to the source within the FWHM of the synthesized beams at 600 MHz is shown in black. Downward triangle markers indicate arrival times for detections with CHIME/FRB or EVN with the number of detections in each active phase indicated above these markers while the upward triangle marker indicates the epoch of non-detection of the source by the Effelsberg telescope. The grey shaded regions show a $\pm2.7$-day interval around estimated epochs of source activity. 
	  We truncate our reported exposure on September 30th, 2019 due to an upgrade of the software (see Methods).  \textbf{b}:  The variation in the daily relative RMS noise at the position of FRB 180916.J0158+65, depicted as coloured points with \PC{1$\sigma$} uncertainties, measured using a collection of pulsars detected by CHIME/FRB. 
 There is clearly substantial exposure and nominal sensitivity to the source well outside the regions of observed activity.
  }
   \label{fig:exposure}
\end{figure}

Alternative search methods, such as folding the events and evaluating the resulting profiles with multinomial distribution, using the H-test\cite{89Jager_htest}, and discrete Fourier transform searches with incoherent harmonic summing\cite{rem02}, also return the 16.35-day period, with nominal significances between 4 and 13$\sigma$ under various assumptions (see Methods).
Figure~\ref{fig:exposure} shows the arrival times of the bursts from FRB 180916.J0158+65 from August 28th, 2018 to February 4th, 2020, together with the daily exposure to the source and instrument sensitivity. The instrument was operating for the majority of this time interval with nominal sensitivity; however, the bursts are detected only in a narrow interval at the reported periodicity.
\vk{Note that the short regular daily exposure of CHIME leads to a degeneracy between specific periods of the order of hours to a day and the 16.35-day period (see Methods). Nevertheless, the statistical significance of an astrophysical periodicity remains unaffected. A modest amount of data with bursts detected outside the CHIME observing windows can determine the true period.}
%\footnote{If the true period is aliased, non-CHIME telescopes could observe bursts at more rapid repetition rates, and also outside the apparent 5-day active window.}.}
We conclude that this is the first detected periodicity of any kind in an FRB source.

\begin{figure}[htbp]
\begin{center}
\centerline{\resizebox{0.5\textwidth}{!}{\includegraphics{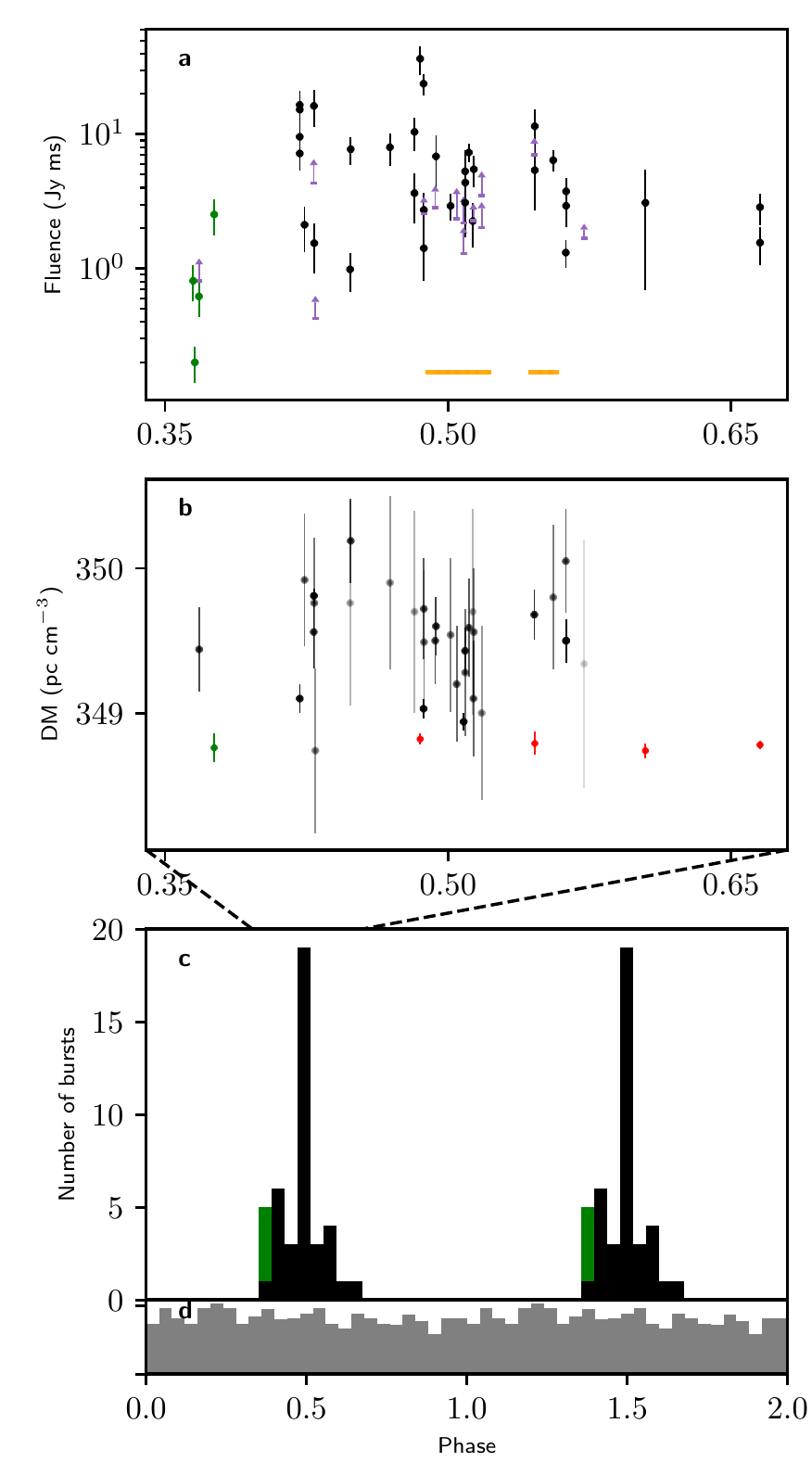}}}
\end{center}
%\end{figure}

%\clearpage
%\addtocounter{figure}{-1}
%\begin{figure} [t!]
	\caption{
\textbf{Burst properties against phase.} FRB 180916.J0158+65 bursts folded at the 16.35-day period with MJD 58369.18 referenced as phase 0. The $1.7$~GHz EVN detection are in green. \textbf{a}: Fluence versus phase. Purple arrows indicate lower bounds derived from CHIME detections; orange lines are the fluence threshold for 1~ms bursts for the search in the $1.4$~GHz Effelsberg data. Error bars on the CHIME detections represent the average fluence 
%variation 
uncertainty
estimated from known point source transits. \textbf{b}: DM versus phase, with 1$\sigma$ error bars. DMs derived from high-resolution baseband data from CHIME/FRB are in red. DMs from CHIME/FRB intensity data are in black; these are subject to potential biases (see Methods). Data points with larger errors are assigned with higher transparency. \textbf{c}: Number of bursts at each phase.
\textbf{d}: Relative exposure time of different phase bins to FRB 180916.J0158+65 within the FWHM of the CHIME synthesized beams at 600 MHz.
}
\label{fig:fluenceDM}
\end{figure}

In addition to the CHIME/FRB detections, FRB 180916.J0158+65 
was detected by the European Very-long-baseline-interferometry Network (EVN)
during 3.5 hours of exposure on June 19th, 2019 
at a central frequency of 1.7 GHz\cite{mnh+20}. 
The EVN arrival times appear at the leading edge of the active phase determined from analysis of CHIME/FRB timing data at lower frequencies of 400$-$800~MHz (Figures~\ref{fig:exposure} and \ref{fig:fluenceDM}).
On the other hand, during a predicted epoch of peak activity (October 29th, 2019 and October 30th, 2019), we observed towards the direction of FRB 180916.J0158+65 for 17.6\,h using the 100-m Effelsberg telescope at 1.4\,GHz (see Methods). In the search, we did not detect any bursts above a fluence limit of 0.17\,{Jy ms} (for bursts with widths of 1\,ms and a detection threshold of S/N$=7\sigma$). These observations were contemporaneous with two CHIME/FRB detected bursts from FRB 180916.J0158+65. 
%The detection of bursts with CHIME/FRB with only $\sim$12 min daily exposure and simultaneous non-detection with Effelsberg with 17.6\,h exposure in two peak active days is suggestive of a frequency dependence to the burst activity. 
Future multi-frequency radio observations of FRB 180916.J0158+65 will aid in understanding the relationship between burst activity and emission frequency.

\PC{We estimate the detection rate of the CHIME/FRB system to be 0.9$^{+0.5}_{-0.4}$ bursts per hour above a fluence threshold of 5.2 Jy ms for a $\pm2.7$-day interval around each epoch of activity.} 
\PC{Additionally, we estimate the detection rate for three sub-intervals in the active phase. We find the rate in a $\pm0.9$-day sub-interval
%(standard deviation in the phases derived by folding burst arrival times) 
centred at the epoch of peak activity to be inconsistent with that estimated for the sub-interval most separated from the epoch of peak activity, suggesting that burst activity is not constant within the active phase (see Methods).}
The 95\% confidence upper limit on the detection rate during inactive phases with non-zero exposure is 0.07 bursts per hour above a fluence threshold of 5.1 Jy ms. 
%Fluence thresholds are complete to the 90\% confidence level.

Figure~\ref{fig:fluenceDM} shows the burst fluence and DM as a function of phase. The phase is computed by folding burst arrival times at a period of 16.35 days with MJD 58369.18 referenced as phase 0. In this definition, phase 0.5 corresponds to the mean of the folded arrival times. There is no apparent trend of burst fluence in the 400--800 MHz band with phase. The best-fit DMs of four CHIME/FRB bursts with high time-resolution data (marked red), as well as the DM of the brightest burst detected by the EVN\cite{mnh+20}, which is accurately determined due to the burst’s narrowness, are consistent with each other, constraining changes in DM to be $<0.1$ pc cm$^{-3}$. These bursts are detected over a time span of $\sim176$ days, which strongly constrains any potential DM variation within
this time period. The DMs of the other bursts, although possibly subject to important biases (see Methods), disfavour DM changes greater than 2 pc cm$^{-3}$ over the full ~400-day span of the
events. In summary, there is no obvious phase- or time-related change in DM in the current data. 

Bursts from FRB 180916.J0158+65 and other repeating FRB sources display complex morphological features\cite{19HesselsRepeaterStructure}: they tend to exhibit 100--200 MHz bandwidth at different central frequencies, with temporal widths of a few ms to tens of ms. Some exhibit downward-drifting sub-bursts at a few to tens of MHz ms$^{-1}$ in the CHIME band\cite{abb+19c}, up to almost a GHz ms$^{-1}$ at 6.5 GHz in the case of FRB 121102\cite{19HesselsRepeaterStructure,jcf+19}. We observe no trend in burst temporal width or bandwidth for FRB 180916.J0158+65 burst detections by CHIME/FRB thus far, neither as a function of time nor phase. Moreover, drifting sub-bursts appear to occur at all phases. In fact, when dedispersed to the average best-fit DM from high-resolution baseband data (as in Extended Data Figure~\ref{fig:bursts}) nearly all bursts seem to exhibit downward-drifting sub-bursts.

\vk{The discovery of a periodicity in a repeating FRB source is an important clue to the nature of this object.
Models in which the emission is 
%expected to be 
purely sporadic (e.g. 
giant pulses as seen from the Crab pulsar\cite{lbp16}, 
or from an isolated young magnetar\cite{lyu14,bel17,metzger2019fast} -- but see below), 
are excluded for this source.
One possible explanation for periodicity is orbital motion, perhaps with the emitter being a neutron star (NS), which could be a radio pulsar or magnetar.  
NSs are commonly found in binary systems with either
stellar/substellar or compact companions.  Massive O/B companions to NSs have been observed, generally in eccentric orbits of many days to months
\cite{jml+92}.
%,kjb+94}.  
The narrow duty
cycle for bursting in the FRB 180916.J0158+65 system, assuming a 16.35-day period,
could be related to eccentricity. The periodicity could reflect phase-related emission (e.g. through an interaction with companion's wind\cite{mz14,zhang2018frb}/asteroid belt\cite{dwwh16}).
The periodicity could also reflect a reduction
in effective radio opacity at apastron, where the radio bursts do not pass through the companion wind/disk\cite{jml+92,mz14}. 
Bursts could also be lensed by companion wind clumps or disk material near periastron\cite{18Main}, but if so,
the preference for downward-drifting frequency structure is puzzling (see Fig.~\ref{fig:bursts}).
With a companion casting material in the orbit, phase-dependent DM or RM variations could eventually be observed, especially with wideband/lower-frequency observations.  Periodic variation could also arise
from classical binary precession, seen in some stellar/NS binaries\cite{kbm+96}.
In Galactic binaries composed of a massive star and pulsar, bright X-ray and $\gamma$-ray emission can be observed\cite{cnv+15}; if detected in this system, such
emission would be a strong diagnostic (but the 149-Mpc distance\cite{mnh+20} here will make a detection challenging).}
A compact object companion to the emitter is also possible.
Given the source's location\cite{mnh+20} in the outskirts of a massive spiral galaxy, a supermassive black hole companion seems unlikely,
however an intermediate- or stellar-mass black hole companion is possible 
and could result in periodically observed bursts if there is relativistic orbital precession.

Isolated compact object scenarios 
%may be consistent with the
%data, but also
may present challenges.  
One popular model to explain repeating FRBs invokes a magnetar central engine \cite{lyu14,bel17,metzger2019fast}.  A periodicity could arise from the rotation of such a star.  However, known Galactic magnetars\cite{ok14} have rotation periods $<$12~s. One young 6.67-hr X-ray pulsar has been argued to be a magnetar\cite{deb+16},
%,tkas17}, 
with torque from a disk of material having spun it down.  However, simulations of such ``fall-back disks'' \cite{xl19} suggest it is difficult to produce periodicities of several hours (the time scale of possible aliased period), let alone several weeks.
A handful of known radio pulsars show intermittent, quasi-periodic emission intervals having periods of several days, and which are known to be magnetospheric in origin\cite{klo+06} (though otherwise poorly understood). However, those sources' radio luminosities are at least nine orders of magnitude lower than in this, and all other, FRB sources.
\vk{Long-term precession in an} isolated spinning pulsar could modulate the visibility \vk{of bright bursts produced through an exotic mechanism. However,}
long-lived precession in isolated neutron stars was 
thought impossible
because of rapid damping due to superfluid vortex line pinning in the
stellar interior \cite{sha77}.  More recent work \cite{alw06,gb19} incorporates modifications to the superfluid properties and may allow for precession.  
If the source were precessing, the bursts would likely be emerging
from a fixed region on the star, likely a magnetic pole, and  the 
$\sim$0.2 burst active phase would suggest either a small viewing angle
or a large polar
region, possibilities that may be distinguishable using
observations of the variation of the position angle of
linearly polarized emission.

\vk{Observations with exposure functions different from that of CHIME are encouraged to test for an aliased periodicity.}
Future observations, both intensity and polarimetric, 
and at all wavebands, could distinguish among models and are strongly encouraged, as are searches for periodicities in other repeaters, to see if the phenomenon is generic.

%\printbibliography[segment=\therefsegment,heading=subbibliography]

\clearpage

%\captionsetup{font={stretch=2}}

\clearpage

\clearpage

\clearpage

\begin{methods}
%\st{SPT: Start with the description of the bursts and the position to which they were barycentered before discussing period finding. Was the barycentering done to the EVN position or to the best fit CHIME/FRB position from the RN1 paper? (won't make a difference for this timescale but we should specify).}

\subsection{Burst characterization:}

Burst DMs and models of dynamic spectra were determined in a manner similar to that used for earlier detections of the source~\cite{abb+19b}. In summary, bursts were dedispersed to a fiducial DM = 349.75 pc cm$^{-3}$ and their DMs were subsequently optimized to maximize structure by calculating the phase coherence of emission in all frequency channels with the \texttt{DM\_phase} package\cite{dmlink}%\footnote{\url{https://github.com/danielemichilli/DM_phase}} 
(Seymour et al. in prep.) over a range of trial DMs. The alignment of sub-bursts in burst dynamic spectra was verified by eye and best-fit values are listed in Extended Data Table~\ref{tab:bursts}.

In addition, we fit structure-optimizing DMs using the same method for four FRB 180916.J0158+65 bursts for which complex voltage (baseband) data were saved to disk: the 181225 and 181226 bursts presented previously\cite{abb+19c}, and the 190604 burst and second 190605 burst presented in this work. The baseband system has a 2.56-$\mu$s time resolution and 0.390625-MHz frequency resolution, but we downsampled the data in time to 40.96, 40.96, 20.48 and 81.92 $\mu$s, for the four bursts respectively, to optimize S/N and resolution. We find DMs 348.78$\pm$0.02, 348.82$\pm$0.02, 348.82$\pm$0.05 and 348.86$\pm$0.05 pc cm$^{-3}$, respectively, best align the sub-structures in the four bursts. We note that DMs measured from CHIME/FRB intensity data, which have 0.98304-ms time resolution, are biased high if a burst is comprised of unresolved downward-drifting sub-bursts that only become obvious at higher time resolution. As the DMs fitted at high time resolution are all consistent with one average DM = 348.82 pc cm$^{-3}$, which is also consistent with the DM of the brightest burst detected by the EVN\cite{mnh+20}, we use that value to dedisperse intensity data for visualization of the millisecond-resolution data.

Dynamic spectra for the four bursts for which the baseband data (at a time resolution of few tens of $\mu$s) were analyzed are shown in Extended Data Figure~\ref{fig:bb_bursts} and dynamic spectra of the intensity data (at $\sim$millisecond time resolution) for the 27 new bursts presented in this work are shown in Extended Data Figure~\ref{fig:bursts}.

Single- or multi-component models of dynamic spectra were fit to millisecond-resolution total-intensity data acquired for each burst using a least-squares algorithm\cite{abb+19a, jcf+19, abb+19b, abb+19c}. We applied two-dimensional models of spectra that consisted of Gaussian temporal shapes and either Gaussian or weighted power-law spectral energy distributions. Due to the complex, varying structure of bursts from this source, we report best-fit parameters using models that do not explicitly fit for one-sided scattering tails, and thus yield estimates of ``observed" widths. Best-fit observed widths and burst arrival times, referenced to the Solar System Barycentre and infinite frequency using per-burst DMs and the EVN position\cite{mnh+20} for this source are presented in Extended Data Table \ref{tab:bursts}.

%One newly reported burst from FRB 180916.J0158+65 -- the second event at MJD 58720} in Extended Data Figure~\ref{fig:bursts}) is comprised of two sub-bursts moving upward in frequency over $\sim 64$ ms (a ``chirp''), but it is unclear whether these are actually sub-bursts under one burst envelope or whether they are two separate bursts. There is no apparent ``bridge'' of emission between the two components.

Three CHIME/FRB events from FRB 180916.J0158+65 (on MJDs 58410, 58720 and 58836) consist of two sub-bursts that occur widely separated in time, with no emission observed between them. In such cases it is unclear whether these pairs of bursts belong to the same envelope emitted by the source, or if they instead represent distinct moments of source activity. Given this ambiguity, we here consider them to be separate bursts. One of the events is reported before\cite{abb+19c}. Therefore, while we report on 10 events registered by the CHIME/FRB instrument previously\cite{abb+19c}, we here consider the event on MJD 58410 to be two bursts, and therefore report values for 11 bursts in Extended Data Table \ref{tab:bursts}.

The burst flux calibration method is nearly identical to that previously presented\cite{abb+19c}, with one crucial difference. Previously we assumed that each burst was detected along the meridian of the primary beam, at the peak sensitivity of its declination arc. Under this assumption, we calibrated using meridian transit observations of bright sources near the declination of each burst to obtain lower-bound fluences and fluxes. For the current analysis, we instead leverage the precise localization\cite{mnh+20} of FRB 180916.J0158.6 combined with our beam model to scale the fluences and fluxes and obtain more accurate values.  From the beam model, we determine a per-frequency scaling between the location of the calibrator at the time of transit and the location of the FRB at the time of burst occurrence. We apply this scaling to the calibrated dynamic spectrum of each burst before generating the band-averaged time series from which the fluence and flux are calculated. If there were multiple sub-bursts in a given burst, then separate fluences and fluxes were obtained for each component. 

We use this method to calculate fluences and peak fluxes for all FRB 180916.J0158.6 bursts detected by CHIME/FRB, including recalculations for the 11 previously published bursts\cite{abb+19c}. More specifically, for each burst located within the FWHM (at 600\,MHz) of the synthesized beam in which it was detected, we complete the scaling analysis using the precise position. For all other bursts, we use the method previously presented\cite{abb+19c} to obtain lower bounds. For each calculation, we use meridian transit observations of the supernova remnant SNR G130.7+03.1, which is located within $1^{\circ}$ in declination from FRB 180916.J0158.6. Fluences and peak fluxes for all bursts are listed in Extended Data Table~\ref{tab:bursts}. For bursts within the beam FWHM, fluences calculated using the precise position are on average $\sim$40\% larger than the corresponding lower bounds calculated assuming detection along the meridian.

Including both newly and previously detected bursts, but omitting those for which only lower bounds could be calculated, we obtain a population of fluences for \bca{25} FRB 180916.J0158.6 bursts, or a total of \bca{33} sub-bursts. The cumulative distribution function of the \bca{33} sub-burst fluences ($N(>F) \propto F^{\alpha + 1}$ where $N$ is the number of bursts detected above a fluence of $F$) is shown in Extended Data Figure~\ref{fig:fluence_distribution}. We determine the power-law index, $\alpha$, of the differential distribution ($dN/dF \propto F^{\alpha}$) using maximum-likelihood estimation methods\cite{csn+09,abp+14}. The turnover in the distribution at lower fluences could be due to our telescope sensitivity limit or the intrinsic burst distribution. To remain agnostic about the cause of the turnover, we pick a fluence threshold that minimizes the Kolmogorov-Smirnov distance between an estimated power law and the underlying distribution. The resulting \bca{5.3}~Jy~ms threshold is shown in Extended Data Figure~\ref{fig:fluence_distribution} as a black dashed vertical line, while our 5.2~Jy~ms active period 90\% confidence completeness threshold is shown as a blue dash-dotted line. Excluding bursts below the Kolmogorov-Smirnov threshold, we complete a Monte Carlo simulation which resamples the fluences according to their uncertainties to obtain a distribution of estimated power law indices determined using a maximum-likelihood estimator at each resampling. The mean of the resulting distribution is \bca{$\alpha = -2.3 \pm 0.3 \pm 0.1$}, where the first error is the statistical uncertainty from the maximum-likelihood estimator and the second error is the standard deviation of the index distribution.

This value is consistent with previous results\cite{gms+19}, which found that the cumulative distribution of energies from FRB 121102 bursts detected at 1.4~GHz using Arecibo followed a power law with $\gamma = -1.8 \pm 0.3$ (equivalent to a differential distribution index of $\alpha = \gamma - 1 = -2.8$). Both of these values are steeper than the $\alpha \sim -1.7$ determined using separate samples of FRB 121102 bursts detected by the Karl G. Jansky Very Large Array at 3~GHz, the Green Bank Telescope at 2~GHz, and the 305-m William E. Gordon Telescope at the Arecibo Observatory at 1.4~GHz\cite{lab+17}. 

Although the comparison of these values is potentially interesting, we reiterate that all are almost certainly contaminated by instrumental biases and likely do not faithfully reflect the underlying source distributions. More robust conclusions about the intrinsic source luminosity function must wait until our CHIME/FRB detection biases are properly evaluated.

\bigskip

\subsection{Statistical significance of the period}

To search for the period, we fold the burst arrival times with different periods, and group the folded bursts into $n$ phase bins \cite{Leahy1983}. 
The folding and grouping of the arrival times make the search relatively insensitive to randomly occurring gaps in the data as long as the exposure is reasonably uniform across phase bins.
 We then perform the classic Pearson $\chi^2$ test for deviation from uniformity: 
\begin{equation}
    \chi^2 =\sum_{i=1}^n \frac{(N_i-E_i)^2}{E_i},
\end{equation} 
where
$n$ is the total number of phase bins, $N_i$ is the observed number of bursts in bin $i$,
$E_i=pT_i$ is the expected number of bursts from a uniform distribution, $T_i$ the exposure time for bin $i$, 
and $p=\sum N_i/\sum T_i$ the average burst rate per unit exposure time. In the case of uniformity, the calculated $\chi^2$ statistic for different folding periods should follow a chi-squared distribution with $n-1$ degrees of freedom (DOF) in the limit as $N \to \infty$. A reduced $\chi^2 \gg 1$ indicates a periodicity unlikely to arise by chance. 
In the study of FRBs, due to the limited sample size, the large $N$ limit may not hold. Then the likelihood of a set of phase bin counts $N_i$ will follow the discrete multinomial distribution. 
\begin{equation}
  P(\{N_i\}) = \frac{N!}{\prod_{i=1}^{n}N_i!} \prod_{i=1}^{n}p_i \end{equation}
where $N=\sum N_i$ is the total number of bursts; $p_i=E_i/N$ is the expected chance to fall into phase bin $i$; and $n$ is again the total number of phase bins. The chance of coincidence with a set of phase bin counts $N_i$ observed corresponds to the sum of the likelihood of all possible permutation $N_k$ that has the same or less likelihood as $P(\{N_i\})$. We verify the significance of any detected periodicity with the chance of coincidence from multinomial distribution. 
%Besides the concerns about the small sample statistics, another problem that would influence the significance is the possibility that the burst arrival times are non-poisson. We will discuss this shortly. 

%Note that this $\chi^2$ test, as well as similar tests, implicitly assumes (1) the arrival time of each burst is independent; (2) the number of bursts in each bin follows a Gaussian distribution. Both of the assumptions may fail in the study of FRB arrival times, where the sample size is small, and the arrival of bursts could be non-Poisson. Therefore, after the search, we have applied several approaches to verify the significance.

We searched for periods from $P_\mr{min}=1.5$ days to $P_\mr{max}=100$ days in steps of $0.1/T$ in frequency (i.e. inverse of the period) space, where $T=506$ days is the longest separation between burst arrival times.
We currently only search for periods of days, as opposed to smaller time scales, to avoid the complexity introduced by the cadence of the source transiting CHIME at a sidereal day and CHIME's beam response during the daily exposure. Moreover, the number of independent periods that need to be searched is proportional to 1/$P$, so the ``look elsewhere'' effect would be orders of magnitude larger for searches with short periodicity, leading to periodicities arising by chance.
%from chances of coincidence.
$P_\mr{min}$ is chosen to be sufficiently larger than a sidereal day such that the modulation of the Fourier spectrum from the sidereal day fades away. $P_\mr{max}$ is chosen so that the extent of the observation time consists of enough cycles. 
We limit our search to a period of $P_\mr{max}=100$ days, 
%so that there could be 
to ensure
at least 5 cycles during the time extent. This is because ``phantom periodicity'' could arise in non-Poisson stochastic process, but %it is
typically only for fewer than 3 cycles\cite{PhantomPeriod16Vaughan}.  
%\rev{A simpler way to estimate this is as follows: we have seen bursts happening in nearby 5 days, therefore, the duty cycle should be greater than 5\% for a period of 100 days.  The chance of coincidence for 5 cycles of burst to fall into 5\% duty cycle would be $0.05^4$ times the number of independent trials ($\sim300$), which is 0.002 ($\sim3\sigma$). This is the lower limit of significance we would start to trust the result. For longer periodicity, we are not confident enough to distinguish periodicity from non-poisson stochastic processes. }
There is a strong instrumental periodicity at a sidereal day due the cadence of the source transiting CHIME, so we avoid searching periods close to the first five subharmonics of a sidereal day $P_\mr{sid}^N=N\,P_\mr{sid}$, with $P_\mr{sid}=0.99727$ day and $N$ is an integer. Given the longest separation between burst arrival times $T$, periods with $|1/P-1/P_\mr{sid}^N|<0.2/T$ are ignored in the search for $N$ between 1 and 5.

After calculating the $\chi^2$ against folding period, we noticed a distinct peak at $P=16.35\pm0.15$ days, and the peak persists with the search using the number of phase bins of 4, 8 and 16. We estimate the significance with 5 phase bins, so that the bin size corresponds to the FWHM of the profile folded at 16.35 day. The reduced $\chi^2=31.7$ with 5 phase bins (4 DOFs). This corresponds to a probability of chance coincidence of $\sim10^{-23}$ (equivalent to 10\,$\sigma$) after taking into consideration the number of independent trials searched $N_\mr{ind}=(1/P_\mr{min}-1/P_\mr{max})T\sim332$. 
 The chance of coincidence from the multinomial distribution is $10^{-20}$, 9.2\,$\sigma$, after multiplying the number of independent trials searched. 
 The period uncertainty is conservatively estimated by $\sigma_P=P\, W_\mr{active}/T_\mr{span}$, where $P$ is the period, $W_\mr{active}$ is the active days in the period, and $T_\mr{span}$ is the longest time separation between burst arrival times. This corresponds to a change in period that 
  would allow the folded 
  pulse
  %phase 
  to drift across the active cycle over the observed time span.
  %, and 
  The calculated error also corresponds to the FWHM of the peak in the periodogram. The period and error are reported at an accuracy of 0.05 day. Within this accuracy, all the following approaches return consistent results.

The above analysis shows that there is a significant $16.35$-day periodicity in the arrival times of FRB 180916.J0158+65 assuming all the bursts are independent events (Poisson statistics). However, for the first repeater, FRB 121102, in which no periodicity has yet
%currently 
been found, the bursts are observed to be clustered\cite{18Oppermann}. This triggers the concern that the arrival times of nearby bursts in FRB 180916.J0158+65 may not be independent. Correlated processes have been shown to contaminate the significance of apparent periodicities in certain contexts\cite{PhantomPeriod16Vaughan}. 
%In order to reduce the influence, 
To mitigate this effect,
we apply the search assuming different correlation lengths. The CHIME daily exposure is only tens of minutes; stochastic processes with correlation timescale smaller than 10 minutes 
should  not introduce a periodicity at tens of days. Also, stochastic processes with correlation timescale longer than 16 days 
cannot
account for 
50\% of the bursts 
appearing
in a 0.6-day window. Therefore, two crucial correlation timescales to consider are 0.6 day, the size of the peak active window, and 5 days, the size of the active phase. 

First, we assume a correlation timescale of 0.6-day for the detected bursts. Then all the bursts detected in the tens of minutes exposure of one sidereal day are correlated events. We assign a total weight of unity to bursts that arrived on the same sidereal day.  After weighting, there are 23 active dates, for the \nbursts{} bursts detected by CHIME/FRB. In this way, we also reduce the influence from inaccurate mapping of beam response within the exposure of a sidereal day.
%Moreover, because of (i) the possibility of having additional clustering of bursts at small time scales, apart from the period of 16 days, and (ii) inaccurate mapping of beam response within the exposure of a sidereal day,  
%we applied a more conservative approach by assigning a total weight of unity to bursts that arrived on the same sidereal day. This is effectively recording the dates when the source is active.
%After weighting, there are 18 active dates, which we refer to as ``effective samples,'' for the \nbursts{} bursts detected by CHIME/FRB.

After weighting the events, the distinct peak at 16.35 days persists in the $\chi^2$ against folding periods, followed by a peak at its harmonic of $\sim 32.7$ days. The reduced $\chi^2$ versus period with 5 phase bins is shown in Figure~\ref{fig:periodogram}a.  
%(for visualization, show linear spacing in folding periods as oppose to frequency).
%(for visualization, we show linear spacing in folding periods as opposed to frequency.)
The corresponding chances of the highest peak from chi-square distribution is $\sim 10^{-11}$, equivalent to 6.8\,$\sigma$. And the significance from the multinomial distribution is 6.1\,$\sigma$. Both approaches have taken into consideration the number of independent trials searched. 
%And the chance of 23 independent samples to falling into the 5-day window is $(5/16.35)^{22}$, which corresponds to 5.9~$\sigma$ after considering the 332 trials searched.} 

 Given that the 16.35-day periodicity is not an integer number of sidereal days, the influence of the exposure map on the significance is small. 
To further exclude the possibility of instrumental periodicity, 
we applied bootstrap tests to 
the mock burst arrival times randomly sampled according to the daily exposure to FRB 180916.J0158+65 within the FWHM of the telescope’s synthesized beams at 600 MHz, 
as well as to a random selection of single bursts from Galactic radio pulsars at similar declinations and detected by CHIME/FRB.  
The mock exposure samples are chosen from the exposure between August 28th, 2018 to September 30th, 2019 ($\sim80$\% of the total time span), since several upgrades were made to the CHIME/FRB detection pipeline after October 2019 making characterization of the exposure and sensitivity variation difficult. 26 out of the total 38 bursts (occurring on 17 of 23 active days and during 10 of 14 active cycles) are detected during this time span and the 16.35-day periodicity is already manifest with this fraction of data. Therefore, if there are instrumental effects leading to the periodicity, we expect it to appear in this mock samples as well.
An example of the reduced $\chi^2$ versus period from two sets of the random samples with the same number of 
active days 
as FRB 180916.J0158+65 is shown in Figure~\ref{fig:periodogram}b (random sample selected according to the exposure map) and Figure~\ref{fig:periodogram}c (random samples from the detection of PSR B2319+60). 
With $10^6$ sets of random samples from the exposure map, and $10^6$ sets each from PSR B2319+60, B0138+59, B2224+65, we have not found any period, unrelated to sidereal day, as significant as the 16.35-day period for FRB 180916.J0158+65. 

%\rev{The FRB 180916.J0158+65 bursts folded at the 16.35-day period are shown in Figure~\ref{fig:fluenceDM}c. All the bursts arrive in a 4-day phase window. Even 

In a conservative approach,
we assume all the bursts arriving in the active phase of the same cycle are clustered due to mechanisms irrelevant to the periodicity. We assigning a total weight of unity for bursts with separation less than 5 days and searching for periods. The 16.35-day periodicity is still the most prominent peak with a significance of 4.5\,$\sigma$ from chi-square distribution, and 4.2\,$\sigma$ from multinomial distribution, after taking into account of the number of trials searched. (Notice that the multinomial distribution deals with discrete events only, so we deal with clustered pulses by repeatedly drawing one random pulse from each clustered group, and we report the average probability of many trials.)

%The chance of 14 cycles of bursts all falling in the 4-day window is only $(5/16.35)^{13}$, which corresponds to $3.8 \sigma$ after taking into account the independent trials searched. 
For any other correlation timescale assumed, its influence on the significance would be less strong than this one. Therefore, we conclude that the periodicity is significant even after taking into account potential clustering. 
 
We also performed the H-test\cite{89Jager_htest} summing over $N$ harmonics with $N$ varying from 1 to 10, and Discrete Fourier Transform searches with incoherent harmonic summing \cite{rem02} on the arrival times of FRB 180916.J0158+65.  With all the approaches, with or without subtracting the exposure map, with different weighting, the $\sim16.35\pm0.15$ day period remains prominent, 
% (with the drift of the center of the peak much less than the error quoted),   
and the significance varies between 5$-$13 $\sigma$. 
No other statistically significant periods (other than those harmonically related to this $\sim16$ day-period) appear in the searched range.

Although the definite value of the significance of the observed periodicity is undetermined due to our incomplete knowledge of the underlying burst distribution, the periodicity is obvious with all the approaches we have tried, and does not exist in any of the control samples. There is no phase evolution against time noticed. Therefore, we conclude that the periodicity of FRB 180916.J0158+65 is significant and astrophysical in origin. 

\bigskip
\subsection{Aliasing due to limited daily exposure}
The short regular daily exposure of CHIME 
could lead to a degeneracy between frequency $f_0=(P_0)^{-1}=(16.35~\mr{day})^{-1}$ and an alias $f_N=N\,f_\mr{sid}\pm f_0$, where $N$ is an integer,
$f_\mr{sid}=(0.99727~\mr{day})^{-1}$ is the frequency of a sidereal day. 
%\rev{This alias is intrinsic to all CHIME-like transiting telescopes.}
%The $f_0=1/16.35$~day$^{-1}$ periodicity could also be the frequency when the active phase of a shorter period is coincident with the CHIME daily exposure window. 
%VK isn't the above just a harmonic, already considered previously?
An upper limit on $N$ can be estimated using observed duty cycle (D) over exposure time 
i.e. the duty cycle of the intrinsic period should 
be longer
than the exposure time. With the observed 16.35-day period, all bursts arrive in 
5-day phase window. With 12~min, i.e. 0.008 sidereal day, daily exposure, the upper limit on $N$ is $5/16.35/0.008=37$. However, $50\%$ of the CHIME bursts are detected in a 0.6-day phase window, with the event rate dropping rapidly towards the edges of the active phase. If 0.6~day is the width of the active phase, the duty cycle is 0.04, and $N > 5$ would be disfavoured. We slightly favour $N=0$ for several reasons: (i) if there is aliasing, the observed frequency $f_0$ should be randomly distributed between 0 and 0.5 day$^{-1}$. 
However, $f_0=(16.35~\mr{day})^{-1}=0.06~\mr{day}^{-1}$ is on the low side of that range, which seems slightly fine-tuned (chances to have $f_0\leqslant0.06$, i.e. p-value is $0.06/0.5= 0.12$);
(ii) if we define the duty cycle D to be the fractional phase that contains half of the events, then $N=0$ has the smallest $D=0.036$, followed by $N=1$, $D=0.044$, with larger D for higher $N$; (iii) for most physical models, we expect the bursts detected in the 1.7~GHz EVN observations to have the same periodicity, with alignment of epochs of activity. Only period 16.35 day ($N=0$), 0.20 day ($N=5$) and 0.14 day ($N=7$) satisfy the condition. But the duty cycles of the latter two periods are twice as wide as the 16.35-day period. 
 Despite those arguments to favour $N=0$, or a periodicity of 16.35-day, we cannot unambiguously rule out other values
of $N$ with current data. Nevertheless, a significant astrophysical periodicity must exist for aliasing to happen. 
The exposure of CHIME, as a transit telescope, is fixed by design, therefore, we cannot rule out aliased periods with CHIME/FRB detections alone. However, a modest amount of exposure in the same band using other telescopes, with bursts detected outside the CHIME observing windows may be able to determine the true period.

\bigskip
\subsection{Exposure \& Fluence Completeness Determination}
The exposure to the EVN position\cite{mnh+20} for the source was calculated by adding up the duration of daily transits across the FWHM region of the synthesized beams of the CHIME/FRB system at 600 MHz. After excluding transits for which the observations were interrupted by pipeline upgrades and testing, we estimate the total exposure to be 64 hours in the interval from August 28th, 2018 to September 30th, 2019. We report exposure up to September 30th 2019 since several upgrades were made to the CHIME/FRB detection pipeline through October 2019 making characterization of the exposure and sensitivity variation difficult. The increased exposure time for several days in February 2019 is due to occurrence of two transits in the same UTC day caused by the differing lengths of a solar and a sidereal day (see Figure \ref{fig:exposure}).

Since the burst activity is not constant within the active phase (see Figure 3c),
%we derive two estimates of the exposure and detection rate for the active phases of the source. 
\PC{we divide the active phase into three sub-intervals to derive estimates of the exposure and detection rate. The first of these sub-intervals is a $\pm$0.9-day 
%period 
\vk{span}
(standard deviation in the phases derived by folding burst arrival times) centred on the epoch of peak activity. The detection rate estimated for this sub-interval is 1.8$^{+1.3}_{-0.8}$ bursts per hour above a fluence threshold of 5.2 Jy ms, based on 13 of the 38 detections. The second sub-interval is defined as the duration between 1$\sigma$ and 2$\sigma$ (i.e. between 0.9 and 1.8 days) of the epoch of peak activity with the third sub-interval starting at 2$\sigma$ and covering the remaining active phase.
%the edge of the $\pm$2.6-day-long active phase. 
The detection rate in the former sub-interval is 0.8$^{+1.0}_{-0.5}$ bursts per hour while that in the latter is 0.1$^{+0.6}_{-0.1}$ bursts per hour, with five and one detections in the two sub-intervals, respectively. The rate in the first sub-interval is inconsistent with that in the third hinting at a variation of burst rate within the active phase. Poissonian uncertainties associated with all rate measurements are at the 95\% confidence level with the fluence thresholds being complete to the 90\% confidence level. } 
%These estimates assume different durations of the active phase, one being the standard deviation in the phases derived by folding burst arrival times at the 16.35-day period ($\pm0.8$-day interval) and the other being a $\pm2.6$-day interval which includes the arrival times of all bursts. 

The variability of the event rate within the active phase can influence the chance of detection in different cycles. 50\% of the bursts are detected in a narrow ~0.6 day phase window. CHIME only has tens of minutes exposure per sidereal day. Therefore, for 40\% of the observed cycle, the 0.6-day peak activity window
lies
outside of the CHIME exposure window, which reduces the chance of detection. This effect will be better studied when more bursts are collected and the event rate 
versus
%against 
phase can be better modelled.

The CHIME/FRB system was operating nominally for a total of 21 hours in the active phases defined as a $\pm2.7$-day interval around each epoch of source activity. Nineteen of the 38 detections occurred during intervals included in this exposure time. A total of 7 hours of the exposure was within a $\pm$0.9-day interval of estimated epochs of peak source activity, with 13 out of the \nbursts{}reported bursts detected in this period. The other 25 bursts were not included in the calculation of the detection rate for various reasons. Six of these bursts were emitted outside the fiducial definition of the active phase, 12 of these bursts were detected after the interval used for the evaluation of the exposure, six were detected when the source location was not within the FWHM region of the synthesized beams at 600 MHz and one was detected on September 16th, 2018 when system metrics were not being recorded by the detection pipeline. \PC{The other two sub-intervals had an exposure of seven hours each.} The exposure in the inactive phase (outside the $\pm2.7$-day intervals around the estimated epochs of activity for the source) was a total of 43 hours.
%, with seven bursts which were excluded from the previous definition of the active phase being included here. 

In order to characterize the variation in sensitivity (due to changes in gain calibration, the detection pipeline and RFI environment) for each day included in the exposure, we use a method described previously\cite{jcf+19}. We analyze the distribution of S/Ns of pulsars detected with the CHIME/FRB system within 10$^\circ$ of the source declination. In contrast to the approach used previously \cite{jcf+19}, we use pulsars which were detected on each sidereal day for which the telescope was operating with the same gain calibration, instead of within a UTC day. For each pulsar, the rms noise is measured relative to the median over all days the pulsar was detected. A weighted average of these measurements for all pulsars provides an estimate of the overall variation in rms noise on each sidereal day.

Fluence completeness was determined following previously reported methods\cite{jcf+19, abb+19c}. {\AJ These methods consider sensitivity variation due to observing epoch, position along transit, and burst spectral shape. In a Monte Carlo simulation with $10^6$ realizations, fluence detection thresholds for different scenarios are generated for a given period of exposure. These simulations produce a set of relative sensitivities, which are tied to a flux scale using bandpass-calibrated observations. Due to the structured nature of both bandpass and transit sensitivity, the final distribution of fluence thresholds is non-Gaussian. Therefore, we take percentiles of the distribution to quote completeness at a given confidence level.} We use the bursts reported in Extended Data Table~\ref{tab:bursts} as reference observations to simulate fluence thresholds for both the active and inactive periods, finding completeness at the 90\% confidence interval of 5.2~Jy~ms and 5.1~Jy~ms, respectively.

\bigskip
\subsection{Effelsberg observations and analysis}

Motivated by the detection of the 16.35-day period, we observed FRB~180916.J0158+65 at 1.4\,GHz using the 100-m Effelsberg radio telescope and PSRIX data recorder\cite{lazarus2016} for 17.6\,hr during a predicted active period (October 29th 2019 and October 30th 2019). Details of the scan start times and durations are shown in Extended Data Table~\ref{tab:effobs}. We observed with a total bandwidth of 250\,MHz, from 1234\,MHz to 1484\,MHz, divided into 2048 channels. The data were recorded with a time and frequency resolution of 131.072\,{$\upmu$s} and 0.122\,MHz, respectively.

The high-time resolution data were analyzed in the search for millisecond-duration radio bursts, using tools from the {\tt PRESTO} suite of pulsar software\cite{ransom2001thesis}. The data were first searched for radio frequency interference (RFI), and the contaminated frequency channels and time intervals were masked using the tool {\tt rfifind}. Dedispersed time series were generated for the DM range 300--400\,{pc\ cm$^{-3}$} in DM steps of 0.3\,{pc\ cm$^{-3}$} using the tool {\tt prepdata}. We searched each dedispersed time series for single pulses using {\tt single\_pulse\_search.py}. The single pulses identified were filtered for RFI in the search for astrophysical bursts, above a 7$\sigma$ threshold, using an automated classifier\cite{michilli2018b,michilli2018c}. Given the frequency resolution of the PSRIX data, the intra-channel smearing is $\sim$ 0.15\,ms at the DM of FRB~180916.J0158+65. This is comparable to the time resolution of the data. Our search was sensitive up to burst widths of 39.3\,ms. Bursts from FRB~180916.J0158+65 have been observed with components as narrow as $\sim$~60\,{$\upmu$s} \cite{mnh+20}. Therefore, it is possible that FRB~180916.J0158+65 is producing weak, narrow bursts that we were not sensitive to in this search.

In this search we did not detect any bursts from FRB~180916.J0158+65 above a 7$\sigma$ threshold. By taking typical values of the system temperature and gain for Effelsberg (T$_{\text{sys}}\approx$ 20\,{K},  $\text{G} \approx$ 1.54\,{K Jy$^{-1}$}), estimating a background temperature of 5\,K using the 408\,MHz all-sky map \cite{remazeilles2015} and extrapolating to 1.4\,GHz using a spectral index\cite{reich1988} of $-$2.7, we use the radiometer equation to derive the fluence limit of our search (following \cite{cordes2003}). Assuming a burst width of 1~ms and a minimum signal-to-noise ratio of 7, we were sensitive to bursts from FRB~180916.J0158+65 above a fluence threshold of 0.17\,{Jy ms}.

Interestingly, during this predicted active epoch of FRB~180916.J0158+65, two bursts, with a fluence exceeding 2\,{Jy\ ms}, were detected by CHIME/FRB (Extended Data Table~\ref{tab:bursts}). The time of arrival of both bursts are within the same Effelsberg scan indicated in Extended Data Table~\ref{tab:effobs}. Given our detection threshold of 0.17\,{Jy ms}, if these bursts were equally bright at 1.4\,GHz, our search was sufficiently sensitive to detect them. The spectra of the two CHIME/FRB bursts peak in the lower half of the 400--800\,MHz band. The detection of bursts with CHIME/FRB and contemporaneous non-detection with Effelsberg suggests that observed activity from FRB~180916.J0158+65 depends on frequency. In addition, the previously detected bursts from FRB~180916.J0158+65 with the EVN at 1.7\,GHz are found at the leading edge of the activity cycle observed at 400--800\,MHz (Figure 3). Future multi-frequency observations of FRB~180916.J0158+65 during predicted active periods are crucial in order to quantify this behaviour.

\end{methods}

\clearpage
%\captionsetup{font={stretch=2}}

\begin{extended-data}
\vspace{-1cm}

\begin{figure}[pthb]
\includegraphics[width=\textwidth]{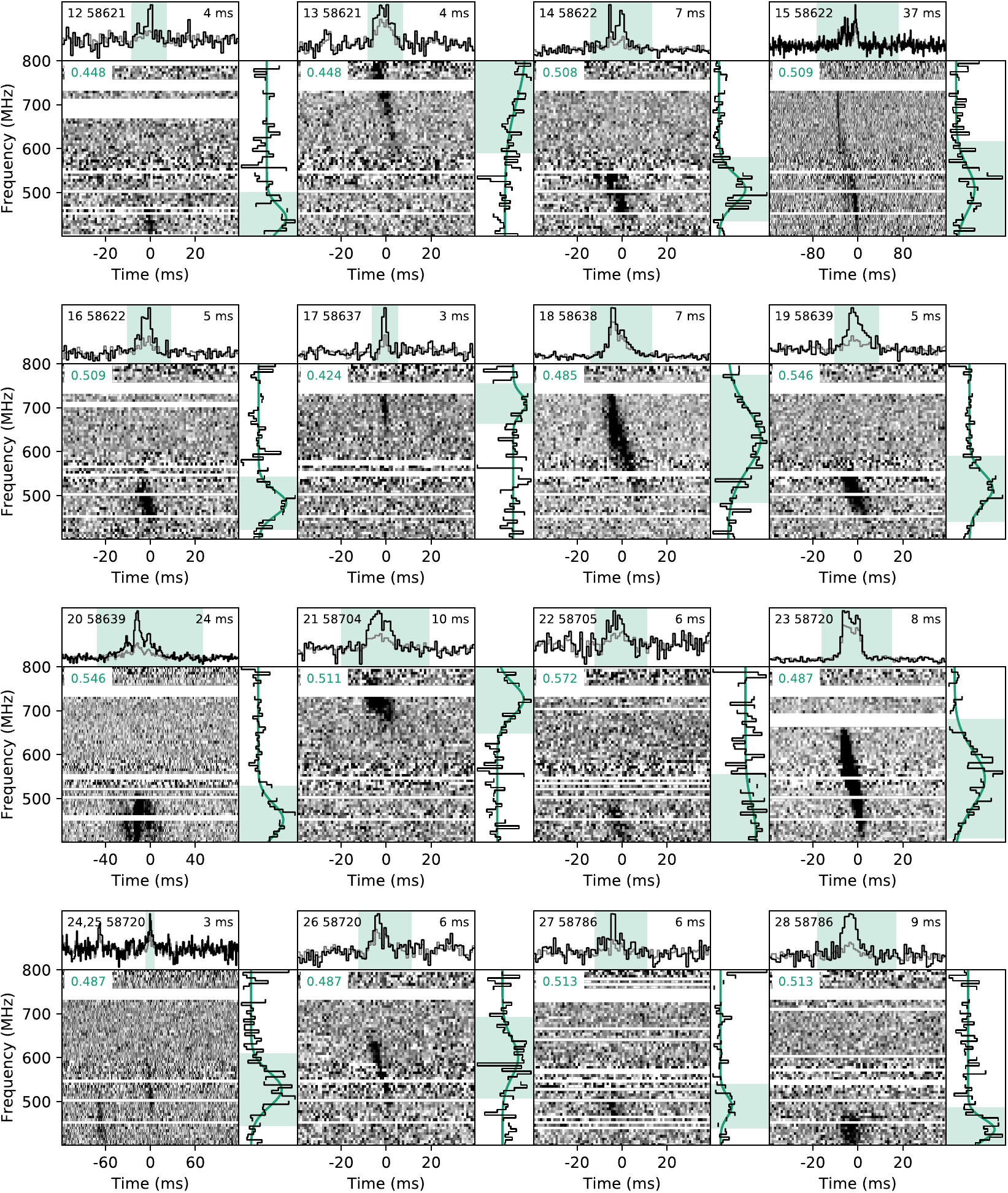}
\end{figure}
\clearpage
\begin{figure}[t]
\caption{\textbf{Dynamic spectra of the newly reported bursts (bursts 12-28 in the Extended Data Table~\ref{tab:bursts})} 
Bursts with separations of less than 0.1 s are shown in joint spectra (bursts 24 and 25). All bursts are dedispersed to 348.82 pc cm−3. The sequence number of the bursts and arrival date (in MJD) are given in the top left coner. The arrival phase of the bursts within the 16.35-d cycle is given in the top left corner of each spectrum. Each plot also gives the 0.98304-ms time-resolution dedispersed intensity data with the integrated burst profile on top and the on-pulse spectrum on the right. Intensity values are saturated at the 5th and 95th percentiles. Pulse widths—defined as the width of the boxcar with the highest S/N after convolution with the burst profile—are given in the top-right corner. The shaded region in each profile (four times the pulse width) was used for the extraction of the on-pulse spectrum. The shaded region in the on-pulse spectrum shows the full width at tenth maximum of a Gaussian fit. In each burst profile, the black line is the integration over the full width at tenth maximum of the spectrum and the grey line is the integration over the full bandwidth. For better visualization, we downsampled the full-resolution data (16,384 channels) to 64 sub-bands, each with a bandwidth of 6.25 MHz. Horizontal white bands represent missing or masked data. There are underlying missing or masked channels at full resolution, resulting in an average effective bandwidth of 224 MHz.
}
\label{fig:bursts}
\end{figure}

\begin{figure}[hbp]
\includegraphics[width=\textwidth]{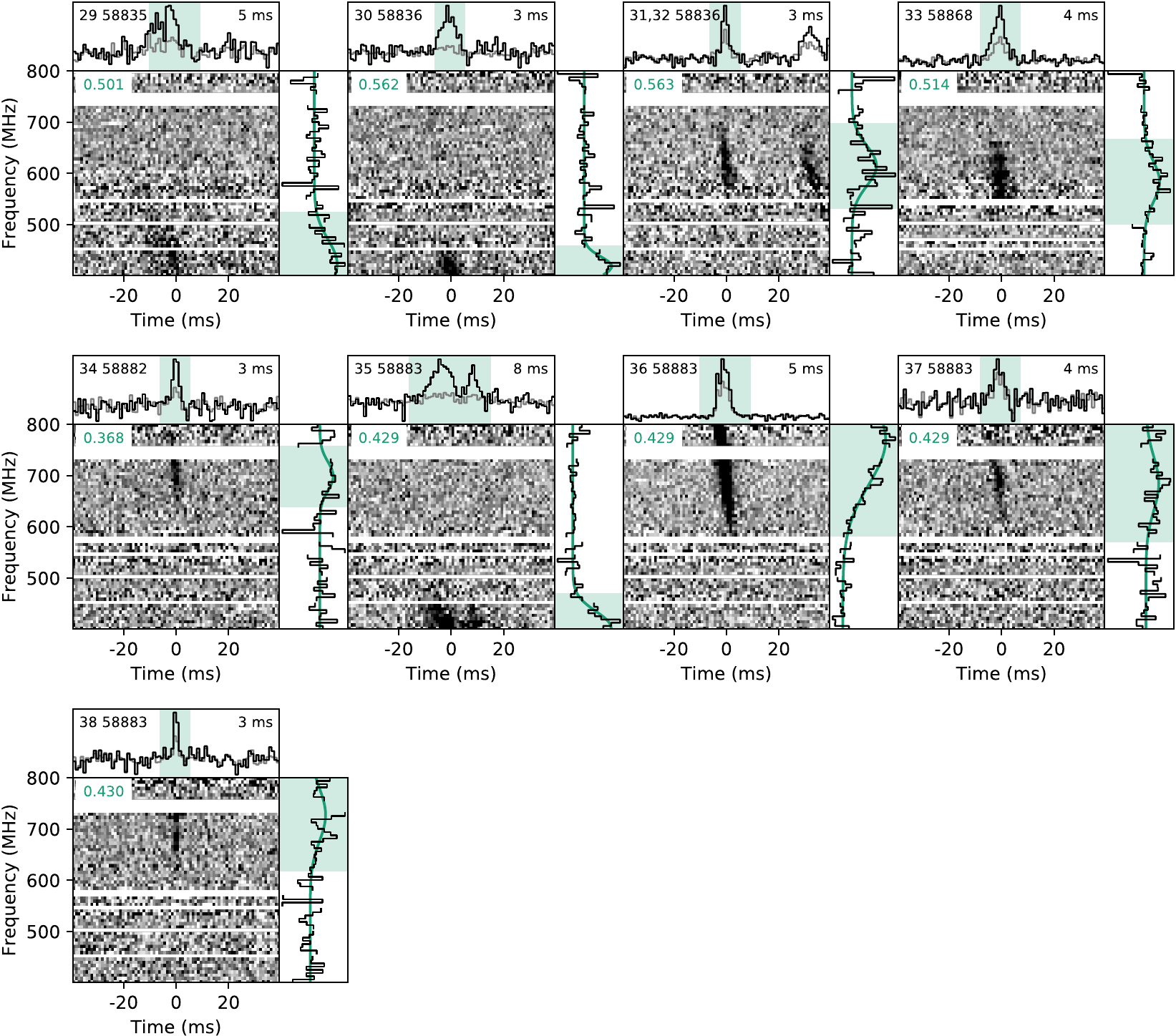}
\caption{\textbf{Dynamic spectra of the newly reported bursts (bursts 29-38 in the Extended Data Table 1).} As in Extended Data Fig.~\ref{fig:bursts}. Bursts 31 and 32 are shown in joint spectra.
}
\label{fig:bursts2}
\end{figure}

\begin{figure}[phb]
\includegraphics[width=\textwidth]{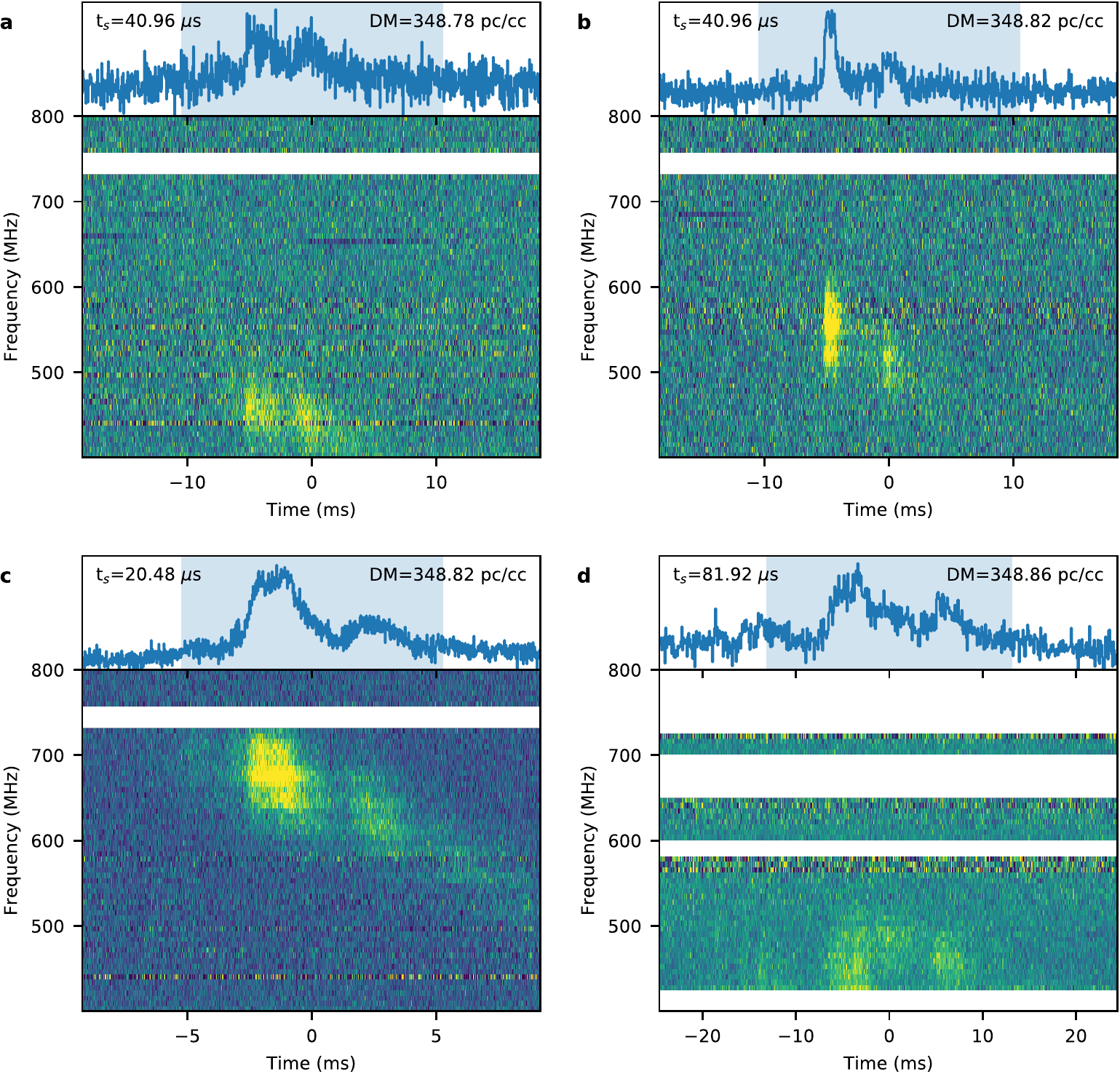}
\caption{
\textbf{Dynamic spectra of bursts with available baseband data.} The 181225 (a), 181226 (b), 190604 (c) and 190605 (d) bursts, dedispersed to the per-burst optimal DMs that are listed in the top right corner of each panel. The sampling time after downsampling is listed in the top left corner of each panel. Intensity values are saturated at the 1$^\mathrm{st}$ and 99$^\mathrm{th}$ percentiles. 64 frequency subbands with a 6.25 MHz subband bandwidth are shown for all bursts. Horizontal white bands represent missing or masked data.
\label{fig:bb_bursts}}
\end{figure}

\begin{figure}[hbp]
\includegraphics[width=\textwidth]{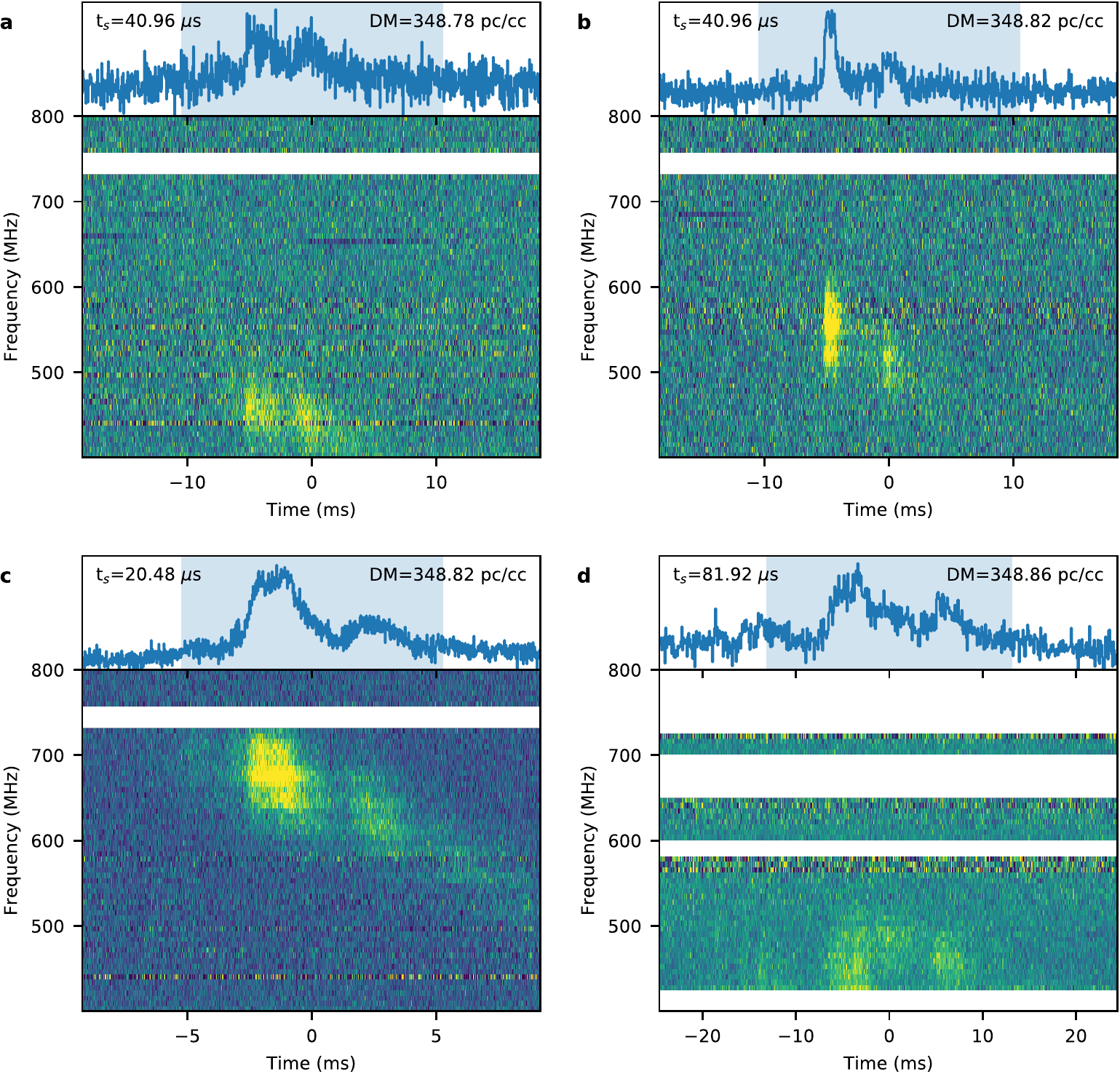}
\caption{
\textbf{Dynamic spectra of bursts with available baseband data.} Burst 9 (a), 10 (b), 18 (c) and 20 (d), dedispersed to the per-burst optimal DMs that are listed in the top right corner of each panel. The sampling time after downsampling is listed in the top left corner of each panel. Intensity values are saturated at the 1$^\mathrm{st}$ and 99$^\mathrm{th}$ percentiles. 64 frequency subbands with a 6.25 MHz subband bandwidth are shown for all bursts. Horizontal white bands represent missing or masked data.
\label{fig:bb_bursts}}
\end{figure}

\begin{figure}[htbp]
\begin{center}
\includegraphics[width=\textwidth]{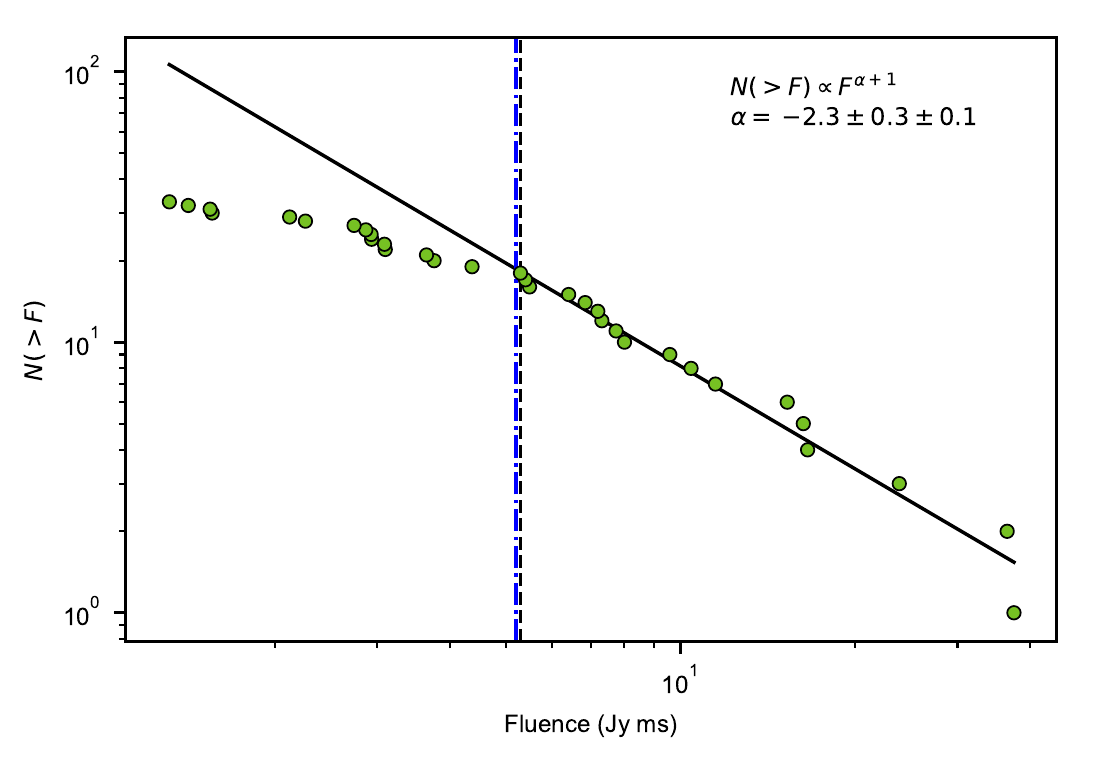}
\caption{{\bf Cumulative distribution of burst fluences.} The distribution is composed of both new bursts and those previously detected\cite{abb+19c}. Excluding bursts detected beyond the 600\,MHz FWHM of any CHIME/FRB synthesized beam, this includes \bca{25} bursts split into sub-bursts, yielding \bca{33} fluence measurements. The black solid line represents the maximum-likelihood estimated power law with differential distribution index \bca{$\alpha = -2.3\pm 0.3 \pm 0.1$}, where the first error is the statistical uncertainty from the maximum-likelihood estimator and the second error is the standard deviation of the distribution of power-law indices obtained from a Monte Carlo simulation that resamples the fluences according to their uncertainties. The black dashed vertical line denotes the \bca{5.3}~Jy~ms threshold determined by minimizing the Kolmogorov-Smirnov distance between a power-law fit and the underlying data. The blue dash-dotted vertical line denotes the 5.2~Jy~ms active period 90\% confidence completeness threshold.}
\label{fig:fluence_distribution}
\end{center}
\end{figure}
\clearpage

\begin{table}[t!]
\caption{
{\bf Burst Properties.} 
Best-fit parameters for \nbursts{} bursts from FRB 180916.J0158$+$65 detected by CHIME/FRB. Uncertainties are reported at the 68.3\% confidence level. Bursts with multiple components have one arrival time and several widths, fluences, and peak fluxes reported; the arrival time refers to the arrival of the first sub-burst from the EVN position at the Solar System Barycentre after correcting to infinite frequency (i.e., after removing the time delay from dispersion) using the listed DM. Fluence and peak flux values for each sub-burst component are presented in order of arrival.
\label{tab:bursts}
}
%\end{table}

%\clearpage
%\addtocounter{figure}{-1}
%\begin{table} [t!]
\scriptsize
\begin{center}
\begin{tabular}{lllccc}
\hline
 & MJD & DM & Total Width & Fluence & Peak Flux \\
 & (pc~cm$^{-3}$) & (ms) & (Jy ms) & (Jy) \\
 \hline
 \multicolumn{6}{c}{Previously Published Bursts$^{4}$} \\
 \hline
1  & 58377.42972096 & 349.2$\pm$0.2 & 1.40$\pm$0.07 & $>$2.3$\pm$1.2 & $>$1.4$\pm$0.6 \\ %event ID 15372046
2  & 58410.34656422$^\ast$ & 349.0$\pm$0.6 & 4.1$\pm$0.3 & $>$3.5$\pm$1.3 & $>$0.6$\pm$0.3 \\ % event ID 18685272 (first burst)
3  & 58410.34656495$^\ast$ & 349.0$\pm$0.6 & 4.4$\pm$0.9 & $>$2.0$\pm$0.8 & $>$0.3$\pm$0.2 \\ % event ID 18685272 (second burst)
4  & 58426.29413444 & 349.5$\pm$0.3 & 1.37$\pm$0.07 & $>$2.8$\pm$0.9 & $>$1.4$\pm$0.5 \\ % event ID 20425523
5  & 58426.30088378 & 349.6$\pm$0.2$^\dagger$ & 6.3$\pm$1.1 & 6.8$\pm$3.0 & 1.0$\pm$0.6 \\ % event ID 20425669
6  & 58442.25174905 & 349.9$\pm$0.6 & 1.10$\pm$0.09 & 8.0$\pm$2.2 & 2.9$\pm$1.1 \\ % event ID 20887865
7  & 58474.17007574 & 349.1$\pm$0.1 & 4.95$\pm$0.4 / 1.51$\pm$0.3 /  & 9.6$\pm$2.6 / 15$\pm$4 / & 1.9$\pm$0.6 / 6.3$\pm$1.8 / \\ % event ID 23491552 \\ 
& & & 3.7$\pm$0.3 / 2.8$\pm$0.3 & 16.5$\pm$4.5 / 7.2$\pm$1.8 & 3.5$\pm$1.0 / 0.9$\pm$0.4 \\
8  & 58475.16454902 & 349.7$\pm$0.7 & 1.67$\pm$0.05 / 6.3$\pm$0.4 & 10.4$\pm$2.9 / 3.6$\pm$1.5 & 1.9$\pm$0.6 / 0.5$\pm$0.3 \\ % event ID 23588210
9  & 58477.16557196 & 348.9$\pm$0.7$^\ddagger$ & 3.8$\pm$0.3 & 3.1$\pm$2.4 & 1.4$\pm$0.8 \\ % event ID 23789294
10 & 58478.15889115 & 348.8$\pm$0.8$^\ddagger$ & 0.87$\pm$0.3 / 3.6$\pm$0.4 & 2.9$\pm$0.8 / 1.6$\pm$0.5 & 1.9$\pm$0.6 / 0.7$\pm$0.3 \\ % event ID 23891929
11 & 58509.06654412 & 349.8$\pm$0.5 & 2.53$\pm$0.13 & 6.4$\pm$1.2 & 1.7$\pm$0.5 \\ % event ID 28293423
 \hline
 \multicolumn{6}{c}{Bursts from This Work} \\
 \hline
12 & 58621.75641235 & 349.8$\pm$0.7$^\dagger$ & 2.5$\pm$0.6 & 1.0$\pm$0.3 & 0.4$\pm$0.2 \\
13 & 58621.76154355 & 350.2$\pm$0.3 & 1.96$\pm$0.16 & 7.7$\pm$1.8 & 1.6$\pm$0.5 \\
14 & 58622.74024356 & 348.9$\pm$0.1 & 0.58$\pm$0.08 / 0.9$\pm$0.1 & $>$1.3$\pm$0.5 / $>$2.2$\pm$0.9 & $>$0.8$\pm$0.4 / $>$0.8$\pm$0.5 \\
15 & 58622.75315853 & 349.4$\pm$0.2 & 8.0$\pm$0.7 / 2.63$\pm$0.16 & 3.1$\pm$1.4 / 5.3$\pm$2.4 & 0.9$\pm$0.5 / 1.0$\pm$0.6 \\
16 & 58622.75441645 & 349.3$\pm$0.4 & 3.6$\pm$0.4 & 4.4$\pm$2.0 & 1.1$\pm$0.7 \\
17 & 58637.71187752 & 349.9$\pm$0.5 & 1.9$\pm$0.2 & 2.1$\pm$0.8 & 1.6$\pm$0.8 \\
18 & 58638.71347350 & 348.82$\pm$0.05$^\ddagger$ & 1.00$\pm$0.05 & 37$\pm$9  & 6.3$\pm$1.9 \\
19 & 58639.70267121 & 349.7$\pm$0.2 & 2.34$\pm$0.08 & $>$7.0$\pm$1.5  & $>$1.3$\pm$0.4 \\
20 & 58639.70713864 & 348.86$\pm$0.5$^\ddagger$ & 3.72$\pm$0.13 / 4.1$\pm$0.4 & 11.5$\pm$4.0 / 5.4$\pm$2.7 & 2.3$\pm$0.7 / 1.1$\pm$0.5 \\
21 & 58704.53530987 & 349.6$\pm$0.3 & 3.43$\pm$0.14 & 7.3$\pm$1.2 & 1.2$\pm$0.3 \\
22 & 58705.53461219 & 349.3$\pm$0.9$^\dagger$ & 4.3$\pm$1.6 & $>$1.7$\pm$0.3 & $>$0.4$\pm$0.2 \\
23 & 58720.49302597 & 349.0$\pm$0.1 & 1.83$\pm$0.03 & 24$\pm$4 & 4.9$\pm$1.0 \\
24 & 58720.49551788$^\ast$ & 349.7$\pm$0.4 & 7.8$\pm$1.3 & 2.8$\pm$0.9 & 0.6$\pm$0.3 \\
25 & 58720.49551860$^\ast$ & 349.7$\pm$0.4 & 5.1$\pm$0.8 & 1.4$\pm$0.6 & 0.5$\pm$0.3 \\
26 & 58720.49669723 & 349.5$\pm$0.5 & 1.3$\pm$0.3 & $>$2.6$\pm$0.5 & $>$0.9$\pm$0.3 \\ 
27 & 58786.31947325 & 349.7$\pm$0.7$^\dagger$ & 3.1$\pm$0.7 & 2.3$\pm$0.8 & 0.9$\pm$0.8 \\ % event ID 60073222
28 & 58786.32497972 & 349.1$\pm$0.4$^\dagger$ & 3.6$\pm$0.4 & $>$2.3$\pm$0.5 & $>$0.5$\pm$0.2 \\ % event ID 60073678
29 & 58835.17721035 & 349.5$\pm$0.5 & 5.0$\pm$0.5 & 2.9$\pm$0.7 & 0.4$\pm$0.2 \\ % event ID 65395717
30 & 58836.17591788 & 350.1$\pm$0.4 & 4.2$\pm$0.4 & 1.3$\pm$0.3 & 0.3$\pm$0.2 \\ % event ID 65511496
31 & 58836.17845766$^\ast$ & 349.5$\pm$0.2 & 1.55$\pm$0.13 & 2.9$\pm$0.9 & 0.9$\pm$0.3 \\ % event ID 65511653 (first burst)
32 & 58836.17845804$^\ast$ & 349.5$\pm$0.2 & 2.9$\pm$0.2 & 3.8$\pm$1.0 & 0.9$\pm$0.3 \\ % event ID 65511653 (second burst)
33 & 58868.07829861 & 349.6$\pm$0.4 & 2.36$\pm$0.10 & 5.5$\pm$1.5 & 1.0$\pm$0.4 \\ % event ID 69509443
34 & 58882.04838586 & 349.4$\pm$0.3 & 1.14$\pm$0.12 & $>$0.8$\pm$0.3 & $>$0.5$\pm$0.2 \\ % event ID 71665813
35 & 58883.04146680 & 349.6$\pm$0.3 & 8.6$\pm$0.5 & $>$4.3$\pm$1.6 & $>$0.4$\pm$0.3 \\ % event ID 71784400
36 & 58883.04307123 & 349.81$\pm$0.05 & 1.157$\pm$0.011 & 16.3$\pm$5.0 & 6.1$\pm$2.0 \\ % event ID 71784510
37 & 58883.04556977 & 349.8$\pm$0.5 & 1.48$\pm$0.13 & 1.5$\pm$0.6 & 0.5$\pm$0.2 \\ % event ID 71784833 
38 & 58883.05523556 & 348.7$\pm$0.6 & 0.76$\pm$0.07 & $>$0.4$\pm$0.1 & $>$0.5$\pm$0.3 \\ \hline % event ID 71785167

\end{tabular}
\end{center}
$^\ast$ Considered to be separate bursts observed during the same CHIME/FRB event due to large time separation and no clear emission between bursts. See Methods for a brief discussion. \\
$^\dagger$ From S/N-optimization.\\
$^\ddagger$ From structure-optimization of CHIME/FRB baseband data.\\
%$^\ddag$ Lower bound(s).
\end{table}

\begin{table}
\caption{\label{tab:effobs} {\bf Effelsberg observations.} Details of the observations of FRB~180916.J0158+65 using the 100-m Effelsberg telescope on 29th October 2019 and 30th October 2019.\\}
\scriptsize
\begin{center}
 \begin{tabular}{l c c c} 
 \hline
 %metric Fluence/$\sqrt{\text{width}}$ [$\mathrm{Jy\ ms}/\sqrt{\mathrm{ms}}$]
{}& {Scan time range (date time UTC)} & {Scan time range$^{\ast}$ (MJD)} & {Duration (s)}    \\ [0.5ex] 
 \hline
 & 2019 Oct 29 22:16:20 -- 2019 Oct 30 00:16:06  & 58785.93233 -- 58786.01550 & 
 7186 \\
 
 & 2019 Oct 30 00:18:40 -- 2019 Oct 30 02:18:23 & 58786.01728 -- 58786.10043 & 
 7183 \\
 
 & 2019 Oct 30 02:20:50 -- 2019 Oct 30 04:20:39 & 58786.10212 -- 58786.18534 & 7190 \\
 
 & 2019 Oct 30 04:23:10 -- 2019 Oct 30 06:22:58 & 58786.18708 -- 58786.27027 & 7188\\ 

 ${\dagger}$& 2019 Oct 30 06:25:30 -- 
 2019 Oct 30 08:25:14
 & 58786.27203 -- 58786.35519 & 7184 \\

 & 2019 Oct 30 08:39:30 -- 2019 Oct 30 10:39:18 & 58786.36509 -- 58786.44829 & 7188 \\ 
 
 & 2019 Oct 30 10:41:50 -- 2019 Oct 30 10:58:53 & 58786.45005 -- 58786.46189 & 1023 \\

 & 2019 Oct 30 19:33:40 -- 2019 Oct 30 21:33:27 & 58786.81938 -- 58786.90257 & 7187 \\

 & 2019 Oct 30 21:36:00 -- 2019 Oct 30 23:35:44 & 58786.90434 -- 58786.98749 & 7184 \\
 
 & 2019 Oct 30 23:38:10 -- 2019 Oct 31 01:00:04 & 58786.98918 -- 58787.04606 & 4914 \\

   \hline 
    \end{tabular}

\end{center}
$^{\ast}$ Times quoted at the Solar System Barycentre after correcting to infinite frequency (i.e. after removing the time delay from dispersion) using a DM\cite{mnh+20} of 348.76 pc cm$^{-3}$.\\
${\dagger}$ Two bursts were detected by CHIME/FRB during this scan.
\end{table}

\end{extended-data}

%\printbibliography[segment=\therefsegment,heading=subbibliography,filter=notother]
\clearpage

%\addbibresource{NewRefs.bib}
%\addbibresource{frbrefs.bib}
%\addbibresource{modpsrrefs.bib}
%\bibliography{NewRefs,frbrefs,modpsrrefs}
\bibliography{main.bbl}

\begin{addendum}
\item[Data and code availability]
 The data and code used in this publication are available at https://chime-frb-open-data.github.io (doi: 10.11570/20.0002) %https://doi.org/10.11570/20.0002
    \item
%    \vk{Thank DRAO, NRC, CFI...}
% Unique acks:
\newcommand{\allacks}{
We thank the Dominion Radio Astrophysical Observatory, operated by the National Research Council Canada, for gracious hospitality and useful expertise. The CHIME/FRB Project is funded by a grant from the Canada Foundation for Innovation 2015 Innovation Fund (Project 33213), as well as by the Provinces of British Columbia and Qu\'ebec, and by the Dunlap Institute for Astronomy and Astrophysics at the University of Toronto. Additional support was provided by the Canadian Institute for Advanced Research (CIFAR), McGill University and the McGill Space Institute via the Trottier Family Foundation, and the University of British Columbia. Research at Perimeter Institute is supported by the Government of Canada through Industry Canada and by the Province of Ontario through the Ministry of Research \& Innovation. The National Radio Astronomy Observatory is a facility of the National Science Foundation (NSF) operated under cooperative agreement by Associated Universities, Inc.
FRB research at UBC is supported by an NSERC Discovery Grant and by the Canadian Institute for Advanced Research. The CHIME/FRB baseband system is funded in part by a CFI John R. Evans Leaders Fund grant to I.H.S.
The Dunlap Institute is funded through an endowment established by the David Dunlap family and the University of Toronto. 
% Swedish Research Council
% Tsar-ing
A.S.H. was partly supported by the Dunlap Institute at the University of Toronto.
B.M. acknowledges support from the Spanish Ministerio de Econom\'ia y Competitividad (MINECO) under grants AYA2016-76012-C3-1-P and MDM-2014-0369 of ICCUB (Unidad de Excelencia ``Mar\'ia de Maeztu'').
B.M.G. acknowledges the support of the Natural Sciences and Engineering Research Council of Canada (NSERC) through grant RGPIN-2015-05948, and of the Canada Research Chairs program.
D.M. is a Banting Fellow.
M.B. is supported by an FRQNT Doctoral Research Award.
M.D. is supported by a Killam Fellowship and receives support from an NSERC Discovery Grant, the Canadian Institute for Advanced Research (CIFAR), and from the FRQNT Centre de Recherche en Astrophysique du Quebec.
P.C. is supported by an FRQNT Doctoral Research Award.
P.S. is a Dunlap Fellow and an NSERC Postdoctoral Fellow.
S.M.R. is a CIFAR Fellow and is supported by the NSF Physics Frontiers Center award 1430284.
U.-L.P. receives support from Ontario Research Fund-Research Excellence Program (ORF-RE), Natural Sciences and Engineering Research Council of Canada (NSERC), Simons Foundation, Thoth Technology Inc, and Alexander von Humboldt Foundation.
V.M.K. holds the Lorne Trottier Chair in Astrophysics \& Cosmology and a Canada Research Chair and receives support from an NSERC Discovery Grant and Herzberg Award, from an R. Howard Webster Foundation Fellowship from the Canadian Institute for Advanced Research (CIFAR), and from the FRQNT Centre de Recherche en Astrophysique du Quebec.
Z.P. is supported by a Schulich Graduate Fellowship. 
F.K. acknowledges support by the Swedish Research Council.
J.W.K is supported by National Science Foundation OIA Award 1458952.
The European VLBI Network is a joint facility of independent European, African, Asian, and North American radio astronomy institutes. Scientific results from EVN data presented in this publication are derived from the following EVN project code: EM135. 
Based in part on observations with the 100-m telescope of the MPIfR (Max-Planck-Institut f\"{u}r Radioastronomie) at Effelsberg. 
We thank
Laura Spitler, Marilyn Cruces and Michael Kramer for their help in
acquiring Effelsberg observing time. 
We thank R. Archibald for discussions regarding the H test. 
We thank Shami Chatterjee for pointing out the aliasing possibility to us.
}

\allacks
    \item[Author Contributions]
           All authors from CHIME/FRB collaboration played either leadership or significant supporting roles in one or more of:  the management, development and construction of the CHIME telescope, the CHIME/FRB instrument and the CHIME/FRB software data pipeline, the commissioning and operations of the CHIME/FRB instrument, the data analysis and preparation of this manuscript.
 	All authors from CHIME collaboration played either leadership or significant supporting roles in the management, development and construction of the CHIME telescope.
 	K.N., J.W.T.H., B.M., Z.P., A.K., F.K. and R.K. performed the analysis of EVN and Effelsberg single-dish data presented here.
    \item[Competing Interests] The authors declare that they have no competing financial interests.
    
    \textbf{Correspondence and requests for
        materials} should be addressed to D. Z. Li  (email: dzli@cita.utoronto.ca).
        
    \textbf{Reprints and permissions information} is available at www.nature.com/reprints
\end{addendum}
\end{document}